\documentclass[final,numberedheadings]{aipproc}

\usepackage{amsmath}
\usepackage{pifont}
\usepackage{revsymb} 
\usepackage{epsf}
\usepackage{psfig}
\usepackage{lscape}
\usepackage{graphicx}
\usepackage{rotating}
\layoutstyle{6x9}
\input amssym.def
\input amssym.tex

\newcommand{\aca}{AcA}
\newcommand{\aj}{AJ}
\newcommand{\apj}{ApJ}
\newcommand{\apjl}{ApJ}
\newcommand{\apjs}{ApJS}
\newcommand{\aap}{A\&A}

\newcommand{\apss}{Ap\&SS}
\newcommand{\araa}{ARA\&A}
\newcommand{\mnras}{MNRAS}
\newcommand{\pasp}{PASP}

\begin{document}

\title{Structure and Evolution of Low-Mass Stars:\\ 
       An Overview and Some Open Problems}

%
%

\author{M. Catelan}{
  address={Pontificia Universidad Cat\'olica de Chile, Departamento de 
       Astronom\'\i a y Astrof\'\i sica, \\ Av. Vicu\~{n}a Mackenna 4860, 
       782-0436 Macul, Santiago, Chile \\ e-mail: {\tt\footnotesize mcatelan@astro.puc.cl}}
}

\begin{abstract}
 A review is presented of some of the ingredients, assumptions and techniques 
 that are used in the computation of the structure and evolution of low-mass 
 stars. Emphasis is placed on several ingredients which are still subject 
 to considerable uncertainty. An overview of the evolution of low-mass 
 stars is also presented, from the cloud collapse phase all the way to the 
 white dwarf cooling curve.  
\end{abstract}

\maketitle


\section{Definition of ``Low-Mass Stars''}
\label{sec:def}

Low-mass stars are self-gravitating gaseous (or, rather, plasmatic) bodies 
that develop electron-degenerate cores (meaning 
that all low-lying energy states are filled and electron pressure 
is accordingly dominated by the Pauli exclusion principle) 
soon after leaving the main sequence 
(MS) phase, and hence undergo the so-called ``helium flash'' (i.e., 
ignition of the triple-$\alpha$ process, whereby three alpha particles are 
converted into a $^{12}$C nucleus, under degenerate conditions) 
at the end of their evolution on the red giant branch (RGB). 
Evolutionary calculations indicate that this corresponds to an upper 
mass limit $M \approx 2-2.5\,M_{\odot}$ (e.g., \cite{asea89,asea90,lg99}). 
Exceptionally, 
and as a consequence of extreme mass loss on the RGB, some such stars may 
directly become helium white dwarfs (WD's) before they are ever able to 
ignite helium in their degenerate cores (e.g., \cite{bdv04} and references 
therein). At the low-mass end, 
on the other hand, one finds that objects below $M \simeq 0.08\,M_{\odot}$ 
are incapable of quiescent hydrogen burning. This corresponds to the commonly  
adopted ``dividing line'' between low-mass stars and the so-called brown 
dwarfs (see, e.g., \cite{cb00}, and references therein). Empirically, the 
latter have recently been associated to the new spectral classes L 
(\cite{jkea99}) and T (\cite{abea99}).

\section{Astrophysical Importance}
\label{sec:import}

Low-mass stars are of great astrophysical importance for a variety of 
reasons, which include the following:  

\begin{itemize}

\item 
Due to the shape of the so-called ``Initial Mass Function'' 
(\cite{es55}; see also \cite{pk02} for a recent review), which gives the 
number of stars that are born in a given population as a function of mass, 
most stars in the 
Universe are low-mass stars. As a consequence, and in spite of their lower 
individual masses, most of the baryonic mass in the Universe currently in 
the form of stars is also contained in low-mass stars (and their remnants), 
even though, to be sure, most baryonic mass actually appears to be in the 
form of gas (\cite{mfea98}). Most importantly, 
low-mass stars dominate the integrated light in old stellar systems, 
including elliptical galaxies and the spheroids (bulges and halos) of 
spiral galaxies. Therefore, in a very real sense, it is not possible to 
understand the properties of distant galaxies without adequate models of 
low-mass structure and evolution. 

\item 
In the same token, (bona-fide) globular 
star clusters are almost exclusively comprised of low-mass stars. Given that  
resolved Galactic globular clusters have long been known (\cite{as53}) to be 
the oldest objects in the Universe for which reliable ages can be determined, 
they play a crucial cosmological role, 
since the oldest stars in the Universe cannot be older than the Universe 
itself (see, e.g., \cite{ogea01,lk03,lk04,kc03,fr04}]. 
The latter, according to the latest results from the WMAP experiment 
\cite{dsea03,dsea07}, is $13.73^{+0.13}_{-0.17}$~Gyr old. 
Cluster ages are determined  
by comparison of the observed color-magnitude diagrams of globular clusters 
with theoretical isochrones,\footnote{An isochrone~-- from the greek 
{\em iso}~= same, {\em chronos}~= age~-- represents the geometric locus
occupied by stars of a given chemical composition that are born with 
different masses. Seemingly, the first ``modern'' isochrones were computed 
by \cite{dl64}. A nice qualitative description of how 
isochrones are built 
is provided in \cite{ii71}, whereas more quantitative details can be 
found, for instance, in \cite{bv92}.} especially around the so-called 
``turn-off point'' (see \S\ref{sec:evolover})~-- which occurs as a 
consequence of hydrogen exhaustion in the core (see, e.g., \cite{mw06} 
for a recent discussion). Recent age 
determinations for the oldest, most metal-poor Galactic globular clusters 
(such as M92~=~NGC~6341) include those by \cite{dvea02} (13.5~Gyr), who also 
revise downward (by incorporating diffusion in their models) the age obtained, 
using high-quality Str\"omgren photometry and a distance-independent method, 
by \cite{fgea00}, which was clearly much higher than the age of the Universe 
favored by WMAP; and those by \cite{asea07} (in the 12-14~Gyr range). While 
these ages appear broadly consistent with the favored 
WMAP age, it should be noted that the recently suggested revision in the 
abundances of elements in the Sun, based on 3D models of the solar 
convective region (e.g., \cite{apea01,maea04,ma05}), will lead, 
according to \cite{lpea07}, to an {\em upward} change (by $\approx 0.7$~Gyr) 
in the cluster ages, which may complicate the agreement between WMAP 
and globular cluster results, especially (e.g., \cite{jc04}) 
when the formation and chemical evolution time before the oldest globular 
clusters were born is properly accounted for. The latter time interval, 
according to \cite{kg05}, is likely to be 
in the $0.3-3$~Gyr range. On the other hand, another recent 
study (\cite{diea06}) concludes instead that the new abundances will indeed 
lead to an increase in cluster ages (by a maximum of $\sim 10\%$), but only 
for {\em open} clusters, globular cluster ages remaining instead essentially 
unchanged with respect to previous models. 

\item 
Low-mass stars can be used as a probe of the properties of fundamental particles
that are already present within the so-called Standard Model of particle physics, 
as well as on the existence and nature of (non-standard) dark matter particle 
candidates (e.g., \cite{gr96,gr00} and references therein). For instance, the 
cooling of low-mass RGB stars would be affected by a non-zero neutrino magnetic 
moment, thereby impacting such important observables as the magnitude of the 
RGB tip, the absolute magnitude of the horizontal branch (HB), the pulsation 
periods of RR Lyrae stars, and the relative proportion of stars on the 
RGB, HB, and asymptotic giant branch (AGB). By comparing theoretical 
models in which a neutrino magnetic moment is included with the 
observations, one obtains a very stringent bound on this quantity 
(\cite{gr90,rw92,mcea96}), which in fact is several orders of magnitude 
more stringent than is possible to achieve at present from laboratory 
experiments (e.g., \cite{gr99} and references therein). Using 
similar arguments, stringent limits can also be placed on the electric charge  
of the neutrino (\cite{gr99}), the mass and other interaction properties of 
the axion (e.g., \cite{rd87,mcea96,gr03,gr07a,gr07b}), and even the size of 
large extra dimensions (\cite{scea02}) -- though in the latter case more 
stringent constraints can be obtained from WD's (\cite{bm02}) and 
(especially) neutron stars (\cite{hr02}). 
 
\item 
Finally, it is also worth recalling that our own existence crucially depends 
on the properties of a low-mass star (the Sun), whose past and future evolution 
are accordingly crucial in determining our origin and destiny in the Universe. 
For instance, the Earth may or may not, depending on the amount of mass lost 
in the form of stellar winds (which unfortunately we are unable to predict 
from first principles at present) during its evolution as a red giant, be 
completely engulfed by the Sun during the latter's evolution as an AGB star, 
when its age is around $12-12.5$~Gyr (\cite{jsea93}). The reason is that, 
while it is clear that the solar radius will expand enormously during its 
evolution as a red giant, the actual orbital radii of the planets will move 
out as a consequence of mass loss from the Sun, the ultimate fate of the 
Earth and the other inner planets depending on how much mass exactly is 
lost during the Sun's evolution as a red giant. In fact, the Earth's oceans 
will likely evaporate and its atmosphere be lost to space even before the 
Sun reaches the turnoff point, when its luminosity becomes 10\% of more 
higher than its present value (see \cite{rc94,gr02} for reviews and 
additional references). 

\end{itemize}

\section{The ``Standard'' or ``Canonical'' Theory of Stellar Structure 
  and Evolution}
\label{sec:std}

While the theory of stellar structure and evolution is one of the oldest 
and most successful of all astrophysical theories, it is still subject to 
a number of approximations and limitations which often make their predictive 
power less optimum than is commonly realized. In the following sections, an 
overview is provided of the main ingredients that enter stellar 
structure/evolution calculations for low-mass stars, highlighting some 
of the current limitations and problems encountered when comparing models 
and observations.  

As an editorial remark, the reader who may be interested in an overview 
of the evolution of low-mass stars before digging into a mathematical 
discussion of the principles of stellar structure and evolution is 
encouraged to proceed directly to \S\ref{sec:evolover}, later returning 
to \S\ref{sec:4BAS} in order to better appreciate the techniques that 
are used to arrive at some of the results discussed in \S\ref{sec:evolover}.

\subsection{The Four Basic Equations of Stellar Structure/Evolution}
\label{sec:4BAS}

The modern phase of stellar structure/evolution studies was born with the 
advent of nuclear astrophysics, when the role of thermonuclear reactions 
in stellar interiors became clear (\cite{hb39,bm39,gg39}). The roughly 70 
years since this major development have been mostly devoted to refining the 
physical ingredients that enter the same basic set of four differential 
equations already known at that time, namely: 

\begin{eqnarray}
 \frac{\partial r}{\partial m} & = & \frac{1}{4\pi r^2 \rho},\label{eq:BAS1}\\
 \frac{\partial P}{\partial m} & = & -\frac{G m}{4\pi r^4},  \label{eq:BAS2}\\
 \frac{\partial L}{\partial m} & = & \epsilon - \epsilon_{\nu} - \epsilon_{g}, 
                                                             \label{eq:BAS3}\\
 \frac{\partial T}{\partial m} & = & -\frac{G m T}{4\pi r^4 P} \nabla, 
                                                             \label{eq:BAS4}
\end{eqnarray} 

\noindent where 

\begin{equation} 
 \nabla \equiv \frac{\partial \ln T}{\partial \ln P}.        \label{eq:BAS5}
\end{equation}

\noindent Here, $r$ is the distance from the star center, and $m$ is the 
mass contained within this distance. $P$, $\rho$, and $T$ are the 
thermodynamic variables pressure, density, and temperature, respectively, 
while $L$ is the luminosity (in units of energy per unit time) at the 
position corresponding to $m$ (or $r$). (When $r$ is used as the independent 
variable, one is talking about the {\em Eulerian formalism}, whereas $m$ is used 
as independent variable~-- as above~-- in the so-called {\em Lagrangean formalism}.)
Finally, $\epsilon$ corresponds to the energy generation rate (in the form 
of thermonuclear reactions), $\epsilon_{\nu}$ to the energy loss rate (in 
the form of neutrinos), and $\epsilon_{g}$ to the work that is performed 
on the gas (per unit time per unit mass) during any expansion/contraction 
of the star (for an ideal gas, it can easily be shown that 
$\epsilon_{g} = T dS/dt$, where $S$ is the specific entropy). 
$G$, as usual, is the Newtonian constant of gravitation. Note 
that partial derivatives have been used throughout to emphasize that the 
physical solutions that satisfy these equations are not stationary, but 
rather evolve with time, as a consequence of the nuclear transmutations in 
the stellar interior and the ensuing changes in chemical composition and 
mean molecular weight $\mu$ brought about by the thermonuclear reactions. 

Each one of these basic equations is extensively discussed in any stellar 
structure/evolution textbook (e.g., \cite{ms58,dc68,cg68,chea04,sc05}), 
so that we will not go through their 
derivation in detail. These equations are, in the order they appear, the 
{\em mass conservation equation}, the {\em hydrostatic equilibrium equation}, 
the {\em energy conservation equation}, and the {\em energy transport 
equation}. While the first two are of chief importance in determining 
the mass profile inside the star (and are in fact the only ones 
used in the theory of self-gravitating, ``polytropic'' gas spheres; see,
e.g., \cite{ae26,sc39}), 
the latter two are of paramount importance in defining the {\em thermal 
profile} of the star. Naturally, the latter can only be accomplished with 
a theory of energy transport inside the star, which must somehow appear 
in relation with eq.~(\ref{eq:BAS4}). We will come back to this point 
later.

\subsection{The Constitutive Equations}
\label{sec:const}

What the reader should realize at this point is that, even though we have 
a system of four equations, we actually have five unknown quantities that 
appear explicitly in them, namely: $r$, $\rho$, $P$, $L$, $T$. Therefore, 
in order to be able to solve the system, we need one additional equation. 
This is given by the {\em equation of state} (EOS), which provides one of 
the thermodynamic quantities in terms of the others; for instance, 
$\rho = \rho(P,T,\mu)$. The reader is referred to the Appendix in 
\cite{dvea00} for a discussion of and references to the different EOS
implementations commonly adopted in low-mass stellar interiors 
work.\footnote{The EOS by A. W. Irwin mentioned in \cite{dvea00} 
and currently in use by several different groups can be found in 
{\tt http://sourceforge.net/projects/freeeos} and 
{\tt ftp://astroftp.phys.uvic.ca/pub/irwin/eos/}.\label{foo:freeeos}}

As such, the EOS is often called one of the 
{\em constitutive equations} of stellar structure/evolution, describing 
as it does a physical property of the matter in the stellar interior. 
The perceptive reader will have noticed that additional constitutive 
equations are still needed in order to solve the problem, since several 
additional quantities (such as the $\epsilon$'s) also enter these equations. 
Within this framework, the complete set of constitutive equations can be 
written as follows: 

\begin{eqnarray}
 \rho & = & \rho(P,T,\mu),                      \label{eq:CON1}\\
 c_P  & = & c_P(P,T,\mu),                       \label{eq:CON2}\\
 \kappa_{\nu} & = & \kappa_{\nu}(P,T,\mu),      \label{eq:CON3}\\
 r_{jk} & = & r_{jk}(P,T,\mu),                  \label{eq:CON4}\\
 \epsilon_{\nu} & = & \epsilon_{\nu}(P,T,\mu),  \label{eq:CON5}
\end{eqnarray} 

\noindent where 
$c_P$ is the specific heat at 
constant pressure, $\kappa_{\nu}$ is the monochromatic opacity, and 
$r_{jk}$ is the thermonuclear reaction rate for reactions that transform 
nuclei of type $j$ into nuclei of type $k$ (the corresponding energy 
generation rate $\epsilon_{jk}$ being given by $Q_{jk} r_{jk}$, where 
$Q_{jk}$ represents 
the mass excess, or the amount of energy released when transforming a 
nucleus of type $j$ into a nucleus of type $k$). $\epsilon_{\nu}$ 
is the energy loss rate due to emission of neutrinos (typically taken 
from \cite{mhea94} or from \cite{niea96}, or from earlier work by the 
Japanese group). The mean molecular weight $\mu$ is given by the following 
expression: 

\begin{equation}
 \mu = \frac{\rho}{m_{\rm amu}}\sum_{i}\left[1+\nu_{e}(i)\right]\frac{X_i}{A_i},  
\label{eq:MU}
\end{equation}

\noindent where $X_i$ indicates the chemical abundance by mass fraction 
of the $i$-th chemical 
element present in the stellar interior (whose atomic mass and number 
are $A_i$ and $Z_i$, respectively), and $\nu_{e}(i)$ is the number of 
free electrons contributed by one average particle of type $i$. 
Under suitable use of the Boltzmann and Saha equations (including, 
as may be indispensable in the stellar interior, the appropriate Coulomb 
corrections), one is able to compute $\nu_{e}(i)$, and hence $\mu$, at 
every point in the interior of a star, as needed for integration of the 
above system of equations. The importance of the Saha equation is that 
it allows one to compute the fractions of each element under different 
ionization stages as a function of the local temperature and density. 

Time evolution in the abundance of element $i$ is then given by the 
following equation: 

\begin{equation} 
 \frac{\partial X_i}{\partial t} = \frac{m_i}{\rho}\left(\sum_{j}r_{ji} - \sum_{k}r_{ik} \right), 
 \label{eq:CHEMEVOL}
\end{equation}

\noindent subject to the constraint $\sum_{i}X_i = 1$. 

Note that it is common practice, in stellar interiors work, to 
characterize the chemical composition of a star using only three numbers, 
namely $X$, $Y$, $Z$~-- which represent the mass fraction of hydrogen, 
helium, and all elements heavier than helium (or ``metals,'' in 
astronomical jargon), respectively. In this case, the proportions 
of the different ``metals'' are usually taken to be the solar ones. In the 
case of the elements formed by successive captures of $\alpha$ particles
(helium nuclei), namely O, Ne, Mg, Si, S, Ca, and Ti, which for spheroids 
are found to present non-solar abundance ratios 
(e.g., \cite{cwea89,mzea06}), it has been shown that 
a simple rescaling of $Z$ that takes into account the $\alpha$-element 
overabundance, according to $Z = Z_0 (0.638\,f_{\alpha} + 0.362)$~-- 
where $Z_0$ is the solar-scaled metallicity, $Z$ is the metallicity 
after account is taken of the $\alpha$-element overabundance, and 
$f_{\alpha}$ is the $\alpha$-element overabundance factor~-- is 
sufficient for the computation of models that look morphologically 
very similar in the $\log L - \log T_{\rm eff}$ diagram (hereafter 
Hertzsprung-Russell diagram, or simply HRD), at least for metallicities 
$Z \lesssim 3\times 10^{-3}$ (\cite{msea93,dvea00}; but see also 
\S\ref{sec:compobs} below for caveats regarding the extension of 
this result to the empirical color-magnitude diagram, hereafter CMD 
for short). 

The preceding discussion shows how eqs.~(\ref{eq:CON1}), (\ref{eq:CON4}),
and (\ref{eq:CON5}) can be used to build a stellar structure model and evolve 
it in time. How do the other constitutive equations enter the picture? In 
order to answer this question, we must first take a look at how energy 
is transported in stellar interiors.

\subsection{Energy Transport in the Stellar Interior}
\label{sec:transp}

In stars, under normal circumstances, there is a steady flow of energy from 
the innermost regions, where thermonuclear reactions take place, to the 
outermost regions -- and, eventually, from the latter to the interstellar 
space. Depending on the thermodynamical properties of the matter, such 
energy transport can occur in three different ways: by {\em radiative 
transfer}, which describes the transport of energy from hotter to cooler 
regions by photons as they interact with matter; by {\em convective 
motions}, which are macroscopic (usually turbulent) 
fluid motions which set in, under 
non-degenerate conditions, when radiative transport is incapable of taking 
care of the energy transport by itself; and by {\em conductive transfer}, 
which becomes very efficient under degenerate conditions but is usually 
quite negligible otherwise. 

\subsubsection{Radiative Transport}
\label{sec:rad}

Under conditions of {\em radiative equilibrium}, when essentially all the 
energy is transported outwards by photons, it can be shown that the 
temperature gradient in eq.~(\ref{eq:BAS5}) takes the form 

\begin{equation} 
 \nabla_{\rm rad} = \frac{3}{16\pi a c G}\frac{\overline{\kappa_{\rm R}} L P}{m T^4}, 
\label{eq:RADGR}
\end{equation}

\noindent where $a$ is the radiation constant, $c$ is the speed of light 
in a vacuum, and $\overline{\kappa_{\rm R}}$ is the so-called {\em Rosseland mean opacity}. 
The latter is computed, on the basis of monochromatic opacities, from the 
following expression: 

\begin{equation} 
 \frac{1}{\overline{\kappa_{\rm R}}} = 
\frac{
\displaystyle\int_{0}^{\infty}\frac{1}{\kappa_{\nu}}\frac{\partial B_{\nu}}{\partial T} d\nu
}
{
\displaystyle\int_{0}^{\infty}\frac{\partial B_{\nu}}{\partial T} d\nu
}.
\label{eq:ROSS}
\end{equation}

\noindent Here $\nu$ is the frequency, and $B_{\nu}$ is the monochromatic 
Planck function. Tabulations of Rosseland mean opacities from the 
OPAL\footnote{{\rm OPAL:} 
{\tt http://www-phys.llnl.gov/Research/OPAL/opal.html}\label{foo:opal}} 
(e.g., \cite{ir96}) and 
OP\footnote{{\rm OP:} {\tt http://vizier.u-strasbg.fr/OP.html}\label{foo:op}} 
(\cite{ms05,nbea05} and reference therein) groups 
are available in the literature and on the groups' respective web 
sites, 
and are almost exclusively employed in state-of-the-art stellar structure 
and evolution models, except for very low temperatures where molecular 
effects not treated in sufficient detail by OP and OPAL become important. 
In the latter case, of particular importance when dealing with stellar 
envelopes and atmospheres, the opacities by Alexander \& Ferguson 
(\cite{af94}) are generally used. We will come back to the subject of 
radiative opacities momentarily.

%
\begin{figure}[t]
  \includegraphics*[width=3in]{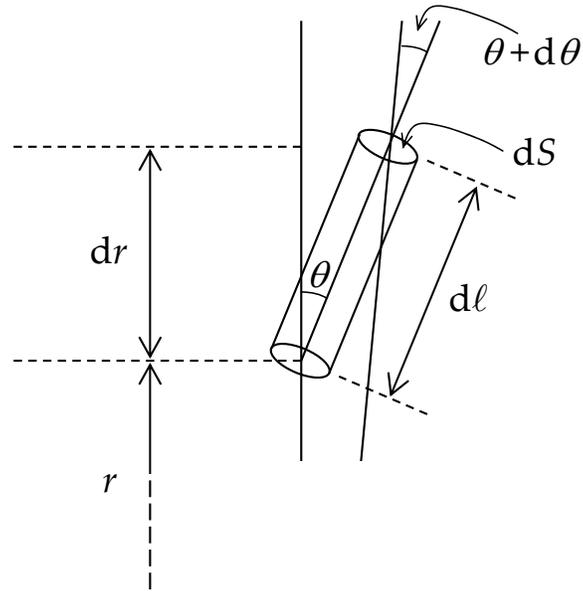}
  \caption{Basic geometric quantities used in the definition of the 
    radiative temperature gradient. Adapted from \cite{ms58}, which 
    in turn is based on the discussion in \S73 in \cite{ae26}. 
   }
      \label{fig:beam}
\end{figure}

Let us first take a look at the radiation transport equation at a given 
point in the stellar interior. 
Consider a point in the star at a distance $r$ from the center. Then the 
amount of energy that flows through a cylinder of length $d\ell$ inclined 
by an angle $\theta$ with respect to the radial vector (Fig.~\ref{fig:beam})
is given by the 
balance among four different terms: first, the radiation that enters 
the base of the cylinder, given by $I(r,\theta)$; second, the 
radiation that is lost from the top of cylinder, given by 
$I(r+dr,\theta+d\theta)$; third, the radiation gains due to 
emission inside the cylinder, given by $dI = j d\ell = 
\textsf{e}\rho d\ell /(4\pi)$, where $j$ is the emission coefficient and 
\textsf{e} is the emissivity (assumed isotropic); fourth, the radiation 
losses due to absorption inside the cylinder, given by 
$dI = - \alpha\, I(r,\theta) d\ell = - \kappa \rho I(r,\theta) d\ell$, 
where $\alpha$ is the 
absorption coefficient and $\kappa$ is the opacity. From this, it follows 
that the {\em energy} that is lost per unit time from inside the cylinder   
due to absorption is $-\kappa \rho d\ell I \, d\Omega \, dS$
(where $S$ is the area of the base, and $\Omega$ the associated solid angle),
whereas the amount of energy gained per unit time due to the emissivity is 
$\textsf{e}\rho d\ell \, dS \, d\Omega /(4\pi)$. Naturally, the difference 
between these two terms must equal 
$[I(r+dr,\theta+d\theta) - I(r,\theta)] d\Omega \, dS$. 
Therefore, one has:

\begin{equation} 
I(r+dr,\theta+d\theta) - I(r,\theta) = \frac{\textsf{e}\rho d\ell}{4\pi}
                                      - I(r,\theta)  \kappa \rho  d\ell,
\label{eq:ALMDIF}
\end{equation}

\noindent which can also be written as 

\begin{equation} 
\frac{\partial I}{\partial r} dr + \frac{\partial I}{\partial \theta} d\theta
  = \frac{\textsf{e}\rho d\ell}{4\pi} - I  \kappa \rho  d\ell. 
\label{eq:ALMDIF2}
\end{equation}

\noindent We must get rid of either $\ell$ or $r$ in order to put this 
equation into a useful form~-- and, of course, we choose to keep the 
latter. Noting the geometrical relations (see Fig.~\ref{fig:beam})

\begin{equation} 
dr = d\ell \cos\theta  \,\,\,\,\,\,\,\,\,\,  {\rm and}   \,\,\,\,\,\,\,\,\,\,  
-d\theta = \frac{d\ell \sin\theta}{r}, 
\label{eq:GEOM}
\end{equation}

\noindent one easily arrives at the following differential equation: 

\begin{equation} 
\frac{\partial I}{\partial r} \cos\theta - 
\frac{\partial I}{\partial \theta}\frac{\sin\theta}{r} 
+ I \kappa \rho - \frac{\textsf{e}\rho}{4\pi} = 0.  
\label{eq:DIFF}
\end{equation}

\noindent Since the only radiation field-related quantity that enters the set 
of eqs.~(\ref{eq:BAS1})-(\ref{eq:BAS4}) is the luminosity, which in turn is 
related to the radiation flux by the geometric relation $F = L/(4\pi r^2)$, 
we can transform the above equation into a more useful form that uses directly 
the flux. We achieve this goal by recalling the definition of flux, namely, 

\begin{equation} 
F(r) \equiv \int_{\Omega} I(r,\theta) \cos\theta d\Omega.  
\label{eq:FLUX}
\end{equation}

\noindent We must accordingly replace the intensity with the flux in 
eq.~(\ref{eq:DIFF}), which we achieve by multiplying the latter by $d\Omega$  
and integrating over all solid angles. We get:  

\begin{equation} 
\frac{\partial}{\partial r} \int_{\Omega} I \cos\theta d\Omega - 
\frac{1}{r}\int_{\Omega}\frac{\partial I}{\partial\theta}\sin\theta d\Omega + 
\kappa\rho \int_{\Omega} I d\Omega - 
\frac{\textsf{e}\rho}{4\pi} \int_{\Omega} d\Omega = 0. 
\label{eq:DIFFF}
\end{equation}

\noindent It is easy to show that, in fact,   

\begin{equation} 
\int_{\Omega}\frac{\partial I}{\partial\theta}\sin\theta d\Omega = -2 F;  
\label{eq:INTOM}
\end{equation}

\noindent recalling the definition of energy density $u(r)$ as the zero-th 
order moment of the radiation field over the speed of light $c$, we then have: 

\begin{equation} 
\frac{\partial F}{\partial r} + \frac{2F}{r} + 
\kappa\rho c u - \textsf{e}\rho  = 0. 
\label{eq:DIFFF2}
\end{equation}

While we have succeeded in getting rid of the intensity and bringing the 
flux into our formalism, we have unfortunately introduced another quantity 
which we (still) do not know how to compute from the thermodynamic properties 
of the medium, namely, the energy density $u$. We thus need at least one 
additional relationship involving the energy density in order to be able 
to solve eq.~(\ref{eq:DIFFF2}). We can obtain one such relation by multiplying 
eq.~(\ref{eq:DIFF}) by $\cos\theta d\Omega$ and again integrating over all solid 
angles. In this way, and recalling that the radiation pressure $P_{\rm R}$ is 
the second moment of the radiation field (over $c$), we get: 

\begin{equation} 
\frac{\partial P_{\rm R}}{\partial r} + \frac{3P_{\rm R}-u}{r} + 
\frac{\kappa\rho F}{c} = 0. 
\label{eq:DIFFPR}
\end{equation}

\noindent Again, we have achieved our goal of eliminating $u$ from the 
differential equation -- but only at the expense of adding in yet another 
unknown quantity, namely the radiation pressure. We could muliply yet 
again eq.~(\ref{eq:DIFF}) by $\cos^{2}\theta d\Omega$ and integrate over all 
solid angles, but this will again add in a higher-order moment of the 
radiation field -- and so on and so forth. 

A particularly elegant way out of this conundrum is to analyze the 
convergence of a cosine series expansion of the radiation field 
at a given point in the stellar interior: 

\begin{equation} 
I(r,\theta) = I_0(r) + \displaystyle\sum_{n=1}^{\infty}I_n(r) \cos^{n}\theta. 
\label{eq:COSEXP}
\end{equation}

\noindent In a typical stellar interior, this series converges extremely 
rapidly (see \S75 in \cite{ae26}, or $\S$II.6 in \cite{ms58}), with a typical 
convergence rate given by 

\begin{equation} 
\left|\frac{I_{n+1}}{I_{n}}\right| \sim 10^{-10},  
\label{eq:COSCONV}
\end{equation}

\noindent which allows us to ignore the higher-order terms in this expansion 
with very high accuracy. Therefore, we take, with sufficient accuracy,  

\begin{equation} 
I(r,\theta) \simeq I_0(r) + I_1(r) \cos\theta. 
\label{eq:COSAPR}
\end{equation}

\noindent In this case, the energy density, the flux, and the radiation 
pressure assume the following values: 

\begin{eqnarray}
 u(r)         & = & \frac{4\pi}{c}I_0(r),                     \label{eq:MOM1}\\
 F(r)         & = & \frac{4\pi}{3}I_1(r),                     \label{eq:MOM2}\\
 P_{\rm R}(r) & = & \frac{4\pi}{3c}I_0(r).                    \label{eq:MOM3}
\end{eqnarray} 

\noindent From this, it is immediately clear that $P_{\rm R} = u/3$. 
It is also apparent that the energy density depends only on the isotropic 
component of the radiation field (higher-order contributions to eq.~[\ref{eq:MOM1}]
typically being many orders of magnitude smaller than the zero-order term, 
and hence totally negligible). Conversely, the radiation flux depends only 
on the degree of anisotropy of the radiation field. Small as the latter may 
be in stellar interiors, it is absolutely crucial in terms of the structure 
of the star and its very existence as a visible object! 

What are the values of $I_0$ and $I_1$? Since $u$ is the energy density  
of the radiation field, and since conditions of thermal equilibrium hold 
to excellent approximation in stellar interiors, it is natural to identify 
$I_0$ with the thermal (Planck) contribution, which is given by 

\begin{equation} 
I_0 = \int_{0}^{\infty}B_{\nu}(T) d\nu = \frac{ac}{4\pi} T^4.   
\label{eq:PLANCK}
\end{equation}

\noindent Therefore, 

\begin{eqnarray}
 u         & = & a T^4,                     \label{eq:T41}\\
 P_{\rm R} & = & \frac{a T^4}{3},           \label{eq:T42}
\end{eqnarray} 
 
\noindent which implies, according to eq.~(\ref{eq:DIFFPR}), 

\begin{equation} 
\frac{\partial P_{\rm R}}{\partial r} + \frac{\kappa\rho F}{c} = 0. 
\label{eq:DIFFPRSIMP}
\end{equation}

\noindent According to eq.~(\ref{eq:T42}), then, 

\begin{equation} 
F = -\frac{4ac}{3\kappa\rho} T^3 \frac{\partial T}{\partial r}, 
\label{eq:FLUXT}
\end{equation}

\noindent or, in terms of the luminosity, 

\begin{equation} 
L = -\frac{16\pi ac r^2}{3\kappa\rho} T^3 \frac{\partial T}{\partial r}. 
\label{eq:LUMT}
\end{equation}

\noindent Recalling the definition of logarithmic temperature gradient  
(eq.~\ref{eq:BAS5}), and bearing in mind the chain rule of calculus, one 
has 

\begin{equation} 
\frac{\partial T}{\partial r} = \frac{\partial T}{\partial P}\frac{\partial P}{\partial r}  
= \frac{T}{P}\frac{\partial\ln T}{\partial\ln P}\frac{\partial P}{\partial r}  
= \frac{T}{P}\frac{\partial P}{\partial r}\nabla_{\rm rad}.  
\label{eq:CHAIN}
\end{equation}

\noindent But we actually know the value of the pressure gradient from our 
hypothesis of hydrostatic equilibrium (eq.~\ref{eq:BAS2}), which in the 
Eulerian form reads $\partial P/\partial r = -Gm\rho/r^2$; therefore, 

\begin{equation} 
\frac{\partial T}{\partial r} = -\frac{Gm\rho T}{r^2 P} \nabla_{\rm rad},  
\label{eq:CHAIN-HYDR}
\end{equation}

\noindent which, together with eq.~(\ref{eq:LUMT}), gives

\begin{equation} 
\nabla_{\rm rad} = \frac{3\kappa P L}{16\pi a c G m T^4}, 
\label{eq:DELRAD}
\end{equation}

\noindent which is the desired expression for the temperature gradient 
in the stellar interior when the energy is transported outwards by 
radiation. This, of course, is the same as eq.~(\ref{eq:RADGR}), showing also 
that the opacity that appears in eq.~(\ref{eq:DELRAD}) is in fact the 
Rosseland mean (eq.~\ref{eq:ROSS}). 

\vskip 0.5cm

\subsubsection{Convective Transport}
\label{sec:conv}

When the temperature gradient indicated by eq.~(\ref{eq:RADGR}) or (\ref{eq:DELRAD})
is too steep, radiation may be unable to carry away all of the energy outwards 
by itself. When this happens, convective instabilities set in, and 
the hypothesis of radiative equilibrium breaks down.\footnote{Very interesting 
discussions of the physical conditions under which this occurs have been 
presented in \cite{ir70,dr93}, where it is argued, on the basis of the 
so-called {\em Naur-Osterbrock criterion} (\cite{rt52,no53}), that it is the 
derivatives of the energy generation rates and radiative opacities with 
$T$ and $\rho$ that define whether a convective core will indeed exist or not.}
In what follows, 
we shall describe the basic theory of convective transport in stellar interiors, 
first establishing the commonly adopted criteria for the onset of the instability, 
and then describing how to compute the value of $\nabla$ in eq.~(\ref{eq:BAS5}) 
using the mixing-length theory of convection. 

\vskip 0.5cm

\noindent{\bf Criteria for the Onset of Convective Instabilities}\label{sec:CRIT}

Let us analyze the conditions under which convective instabilities may set in 
in the stellar interior. With this purpose in mind, we consider the displacement 
of a random fluctuation or ``bubble'' that forms inside the star, and consider 
whether the bubble, after a small displacement, tends to move back to its 
original position (in which case the instability cannot set in) or instead 
tends to keep on moving farther away from its initial position (in which 
case convective instability ensues).

%
\begin{figure}[t]
  \includegraphics*[width=5in]{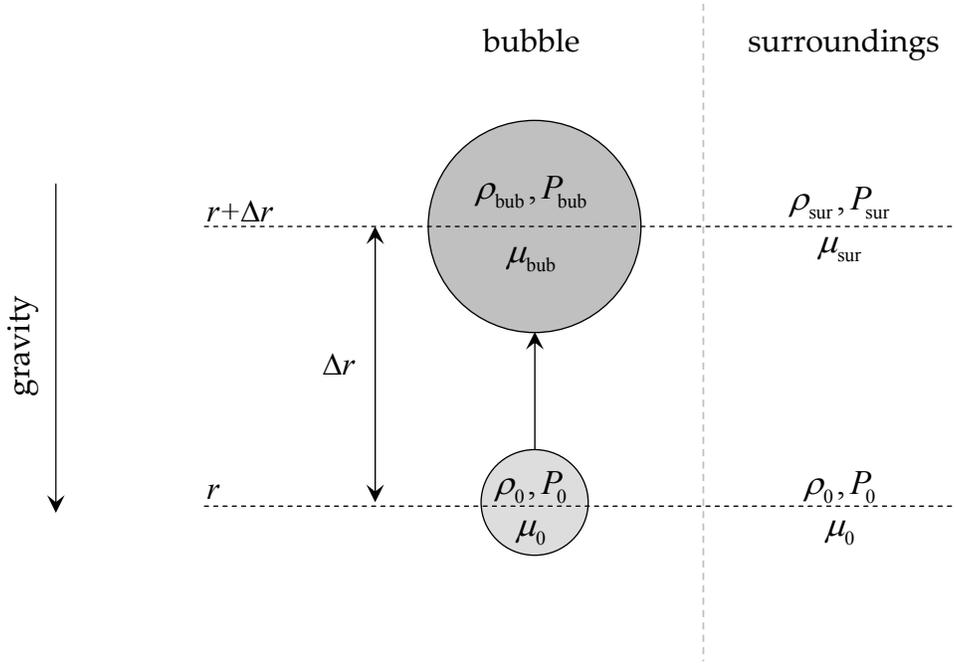}
  \caption{Schematic description of quantities related to the treatment of
    convective instabilities in the stellar interior. In gray is shown a 
    convective bubble element, which is initially in equilibrium with its 
    surroundings (initial conditions are denoted with a subscript ``0'').
    After a displacement by a distance $\Delta r$, the bubble's thermodynamic 
    state will be described by the variables $\rho_{\rm bub}$, 
    $P_{\rm bub}$, $\mu_{\rm bub}$, whereas its surroundings are described 
    instead by $\rho_{\rm sur}$, $P_{\rm sur}$, $\mu_{\rm sur}$. 
   }
      \label{fig:bubble}
\end{figure}

The situation is shown schematically in Fig.~\ref{fig:bubble}. The bubble is 
originally indistinguishable from its surroundings, so that its initial 
thermodynamic properties and those of the surrounding medium are the 
same, and can be characterized by the pressure $P_0$, density $\rho_0$, and 
mean molecular weight $\mu_0$. After a small displacement upwards by a 
distance $\Delta r$, the 
bubble's thermodynamic state is described by the quantities $P_{\rm bub}$,
$\rho_{\rm bub}$, and $\mu_{\rm bub}$, whereas the state of the surroundings 
is described instead by $P_{\rm sur}$, $\rho_{\rm sur}$, $\mu_{\rm sur}$. 
Clearly, by the Archimedes Principle, the bubble will continue moving 
upwards, and hence convective instability set in, if 

\begin{equation} 
\rho_{\rm bub} < \rho_{\rm sur},  
\label{eq:ARCHIM}
\end{equation}

\noindent or, equivalently, 

\begin{equation} 
\frac{d \rho_{\rm bub}}{dr} - \frac{d \rho_{\rm sur}}{dr} < 0.   
\label{eq:ARCHIM-DER}
\end{equation}

\noindent It is easy to see that, as a consequence of the EOS
(eq.~\ref{eq:CON1}), this sets a limit on the rate of variation of the 
temperature gradient inside the star (the pressure gradient is already 
specified by the hydrostatic equilibrium condition, eq.~\ref{eq:BAS2}). 
We can make further 
progress by noting that, provided the speed of sound is sufficiently high 
compared with the displacement velocity of the bubble, the latter is always 
capable of keeping pressure equilibrium with its surroundings. [Indeed, 
according to the recent calculations by \cite{tc06}, supersonic convection 
does not occur in the case of low-mass stars, except perhaps at its high-mass 
end (\S\ref{sec:def}); see his Fig.~1.]
Based on this assumption, let us compute
the requirement on the temperature gradient explicitly (similar discussions 
can be found in \cite{sk66,sc05}). 

First, note that the net acceleration of the bubble can be computed, if 
viscous drag forces can be ignored, from the following expression:  

\begin{equation} 
\frac{d^2 \Delta R}{dt^2} = - g - \frac{1}{\rho}\frac{dP}{dr};   
\label{eq:ARCHIM-FORCE-0}
\end{equation}

\noindent now taking into account the fact that $\Delta r$ is a small 
displacement, one may write $\rho = \rho_0 (1+\Delta\rho/\rho_0)$ 
(where $\Delta\rho \equiv \rho_{\rm bub} - \rho_{\rm sur}$). Therefore, 
to first order, one has

\begin{eqnarray} 
\frac{d^2 \Delta R}{dr^2} & = & - g - 
\frac{1}{\rho_0\left(1+\displaystyle\frac{\Delta\rho}{\rho_0}\right)} 
\frac{dP}{dr} \\
 & \simeq & 
- g_0 - \frac{1}{\rho_0}
\left(\frac{dP}{dr}\right)_0 
\left(1-\frac{\Delta\rho}{\rho_0}\right) \\
 & = & 
- g_0 - \frac{1}{\rho_0}\left(\frac{dP}{dr}\right)_0 + 
\frac{1}{\rho_0}\left(\frac{dP}{dr}\right)_0 \frac{\Delta\rho}{\rho_0}  \\
 & = & 
- g_0 \frac{\Delta\rho}{\rho_0}, 
\label{eq:ARCHIM-FORCE-1}
\end{eqnarray}

\noindent where we have used the fact that the bubble is initially in 
equilibrium, i.e., 

\begin{equation} 
\left(\frac{d^2 \Delta R}{dr^2}\right)_0  =  - g_0 - \frac{1}{\rho_0} \left(\frac{dP}{dr}\right)_0 = 0.
\end{equation} 

Therefore, dropping subscripts, one may write, to first order,  

\begin{equation} 
\frac{d^2 \Delta R}{dr^2} = - g \frac{\Delta\rho}{\rho}.  
\label{eq:ARCHIM-FORCE}
\end{equation}

\noindent We can 
obtain additional differential equations describing the time behavior of 
the thermodynamic properties of our bubble as follows. For the mean 
molecular weight, one has

\begin{equation} 
\Delta\mu \equiv \mu_{\rm bub} - \mu_{\rm sur} = 
\mu_0 - \left(\mu_0 + \frac{d\mu}{dr}\Delta r\right) = -\frac{d\mu}{dr}\Delta r, 
\label{eq:DMU}
\end{equation}

\noindent which, using the chain rule and also considering that the evolutionary 
timescales are always much longer than the convective instability timescales, 
can also be written as 

\begin{equation} 
\frac{d\Delta\mu}{dt} = -\frac{d}{dt}\left(\frac{d\mu}{dr}\Delta r\right)
= -\mu \frac{d\ln\mu}{d\ln P} \frac{d\ln P}{dr}\frac{d\Delta r}{dt}
= \mu \nabla_{\mu} \lambda_P^{-1} \frac{d\Delta r}{dt},
\label{eq:DIFF-DMU}
\end{equation}
 
\noindent which is the desired equation for the perturbation in $\mu$, and 
where we have defined $\nabla_{\mu} \equiv d\ln\mu/d\ln P$ and the {\em pressure 
scale height} $\lambda_P^{-1} \equiv -P^{-1}dP/dr$. Likewise,
for $T$, we define 

\begin{equation} 
\Delta T \equiv T_{\rm bub} - T_{\rm sur} = 
\left[\left(\frac{dT}{dr}\right)_{\rm ad} - \left(\frac{dT}{dr}\right)_{\rm rad}\right]\Delta r
- \beta \Delta T dt. 
\label{eq:DMT}
\end{equation}

\noindent In this equation, the temperature gradient that would be present if 
the displacement of the bubble were precisely adiabatic (i.e., if the bubble did 
not exchange any heat with its surroundings) is denoted by $(dT/dr)_{\rm ad}$;
since Nature is never precisely adiabatic, we have also taken into account 
non-adiabatic effects by adding in the second term on the right, where one 
can see that $\beta\Delta T$ is the rate of temperature change due to the 
bubble's energy losses. Obviously, the adiabatic case corresponds to $\beta = 0$. 

Taking the time derivative of eq.~(\ref{eq:DMT}), and again applying the chain 
rule, one gets 

\begin{equation} 
\frac{d\Delta T}{dt} = -T \lambda_P^{-1}
\left[\left(\frac{d\ln T}{d\ln P}\right)_{\rm ad} - \left(\frac{d\ln T}{d\ln P}\right)_{\rm rad}\right]\frac{d\Delta r}{dt}
- \beta \Delta T; 
\label{eq:DMT2}
\end{equation}

\noindent recalling the definition of $\nabla$ (eq.~\ref{eq:BAS5}), one then 
finds 

\begin{equation} 
\frac{d\Delta T}{dt} = 
- T \lambda_P^{-1} \left(\nabla_{\rm ad} - \nabla_{\rm rad}\right)\frac{d\Delta r}{dt}
- \beta \Delta T. 
\label{eq:DMT3}
\end{equation}

We now have a system of three differential equations describing the time evolution 
of the four unknowns $\Delta r$, $\Delta \rho$, $\Delta \mu$, $\Delta T$. 
A fourth relation is provided by the EOS, which dictates that the changes in 
the thermodynamical properties of a system must be related as 

\begin{equation} 
- \frac{\Delta P}{P} + 
\chi_{\rho}\frac{\Delta\rho}{\rho} + \chi_{T}\frac{\Delta T}{T} + 
\chi_{\mu}\frac{\Delta\mu}{\mu} = 0, 
\label{eq:DMEOS0}
\end{equation}

\noindent where 

\begin{equation} 
\chi_{\rho} \equiv \left(\frac{d\ln P}{d\ln\rho}\right)_{T,\mu}, \,\,\,\,\,\,\,\,
\chi_{T}    \equiv \left(\frac{d\ln P}{d\ln T}\right)_{\rho,\mu},\,\,\,\,\,\,\,\,
\chi_{\mu}  \equiv \left(\frac{d\ln P}{d\ln\mu}\right)_{\rho,T}. 
\label{eq:CHI}
\end{equation}

\noindent For instance, for an ideal gas EOS, one can easily see that 
$\chi_{\rho} = \chi_{T} = -\chi_{\mu} = 1$. Since the bubble always 
keeps pressure equilibrium with its surroudings, eq.~(\ref{eq:DMEOS0}) 
simplifies to 

\begin{equation} 
\chi_{\rho}\frac{\Delta\rho}{\rho} + \chi_{T}\frac{\Delta T}{T} + 
\chi_{\mu}\frac{\Delta\mu}{\mu} = 0,  
\label{eq:DMEOS}
\end{equation}

\noindent which is the fourth equation that we were seeking. 

Our system of four equations given by eqs.~(\ref{eq:ARCHIM-FORCE}),
(\ref{eq:DIFF-DMU}), (\ref{eq:DMT3}), and (\ref{eq:DMEOS})
is in fact quite complex, so instead of studying the full motion of the convective 
bubble we simply check whether solutions of the form 

\begin{equation}
\Delta y = A_y {\rm e}^{\nu t},  
\label{eq:FORM}
\end{equation}

\noindent where $A_y$ is a constant and $\nu$ is a frequency, are possible. Inserting this 
proposed solution into the aforementioned equations, we get: 

\begin{eqnarray}
 g A_{\rho} {\rm e}^{\nu t} + \rho\nu^2 A_r {\rm e}^{\nu t} & = & 0,                    \\
 \nu A_{\mu} {\rm e}^{\nu t} - \mu\nabla_{\mu}\lambda_P^{-1}\nu A_r {\rm e}^{\nu t} & = & 0, \\
 \nu A_{T} {\rm e}^{\nu t} - T \lambda_P^{-1} \left(\nabla_{\rm rad}-\nabla_{\rm ad}\right)\nu A_r {\rm e}^{\nu t} 
    + \beta A_T {\rm e}^{\nu t} & = & 0, \\
 \chi_{\rho}\frac{A_{\rho}}{\rho} + \chi_{T}\frac{A_T}{T} + \chi_{\mu}\frac{A_{\mu}}{\mu} & = & 0. 
\end{eqnarray} 

\noindent Now, in order to ensure that a non-trivial 
solution exists, the determinant formed by the $A_x$ coefficients must be zero;
thus:  

\begin{equation}
 \begin{vmatrix}
   0             &  \displaystyle g    &   0   &    \rho\nu^2                                                  \\
&&&\\
   0             &   0    &  \displaystyle\nu  &  \displaystyle -\nu\mu\lambda_P^{-1}\nabla_{\mu}          \\
&&&\\
\displaystyle \displaystyle\nu+\beta       &   0    &   0   &  \displaystyle -\nu T \lambda_P^{-1}\left(\nabla_{\rm rad}-\nabla_{\rm ad}\right)  \\
&&&\\
\displaystyle\frac{\chi_T}{T} & \displaystyle\frac{\chi_{\rho}}{\rho} & \displaystyle\frac{\chi_{\mu}}{\mu} & 0 
 \end{vmatrix} = 0.
\label{eq:DETERM}
\end{equation}

\noindent This leads to the following dispersion relation: 

\begin{equation} 
\nu^3 + \nu^2 \beta - 
\nu \left[g\lambda_P^{-1}\frac{\chi_T}{\chi_{\rho}}\left(\nabla_{\rm rad}-\nabla_{\rm ad}+\frac{\chi_{\mu}}{\chi_T}\nabla_{\mu}\right)\right]
-g \lambda_P^{-1} \beta \frac{\chi_{\mu}}{\chi_{\rho}}\nabla_{\mu} = 0.
\label{eq:DISPER}
\end{equation}

In particular, in the adiabatic case ($\beta = 0$), this reduces to 

\begin{equation} 
\nu^2 = 
g\lambda_P^{-1}\frac{\chi_T}{\chi_{\rho}}\left(\nabla_{\rm rad}-\nabla_{\rm ad}+\frac{\chi_{\mu}}{\chi_T}\nabla_{\mu}\right), 
\label{eq:DISPER-AD}
\end{equation}

\noindent which corresponds to the so-called {\em Brunt-V\"ais\"al\"a} (or buoyancy) 
frequency, first derived in the context of oscillations of planetary atmospheres by 
\cite{vv25} and, independently, by \cite{db27}. 
In this sense, this quantity plays a particularly important role in the theory of 
non-radial oscillations of stars: note that, for $\nu^2 < 0$, the Brunt-V\"ais\"al\"a 
frequency is purely imaginary, and eq.~(\ref{eq:FORM}) indicates a purely oscillatory 
motion -- which is related to the $g$-modes of non-radial pulsations (see, e.g., 
\cite{gs95} and references therein). We are not chiefly preoccupied with oscillations 
of stars though; our interest is focused instead on the possibility of real and 
{\em positive} values of $\nu$, since eq.~(\ref{eq:FORM}) indicates to us that, in 
this case, our initially very small perturbations will actually grow exponentially 
over time, thus giving rise to convective instability. Therefore, in order for 
convective motions to set in in the adiabatic case, one needs 

\begin{equation} 
\nabla_{\rm rad} > \nabla_{\rm ad}-\frac{\chi_{\mu}}{\chi_T}\nabla_{\mu}.  
\label{eq:LEDOUX}
\end{equation}

\noindent This is the so-called {\em Ledoux criterion}, first derived by Ledoux in 
1947 (\cite{pl47}). Note that the mean molecular weight gradient that appears in 
this equation must be treated with care; according to \cite{cg68}, in particular, 
molecular weight gradients that are solely due to ionization effects should 
not be taken into account when using this equation, since ionization effects are 
already implicitly taken into account in the EOS. 

In the special case where there is no chemical composition gradient, 
$d\ln\mu/d\ln P = 0$, and the Ledoux criterion reduces to 

\begin{equation} 
\nabla_{\rm rad} > \nabla_{\rm ad},  
\label{eq:SCHWARZ}
\end{equation}

\noindent which is the well-known {\em Schwarzschild criterion} (\cite{ms06})
for the onset of convective instability. This is~-- by far~-- the most applied 
convective instability criterion in actual stellar structure calculations. The 
way it is usually employed is as follows: first, one integrates the set of 
equations~(\ref{eq:BAS1})-(\ref{eq:BAS4}), assuming radiative equilibrium; 
this gives us $\nabla_{\rm rad}$ at each point in the star. In parallel, one 
also computes the adiabatic gradient $\nabla_{\rm ad}$ at each point, and then 
compares the two. If the inequality (\ref{eq:SCHWARZ}) is satisfied, then the 
radiative temperature gradient is incorrect, and a different recipe must be 
used, taking into account convective energy transport, to define the new 
temperature gradient. Note that $\nabla_{\rm ad}$ can be computed from 
the properties of adiabatic transformations; in other words, 

\begin{equation} 
\nabla_{\rm ad} = \left(\frac{d\ln T}{d\ln P}\right)_{\rm ad} = 1 - \frac{1}{\Gamma_2},
\label{eq:AD-GAMMA}
\end{equation}

\noindent where $\Gamma_2$ is the {\em second adiabatic exponent}, which can 
be computed on the basis of the thermodynamic properties of the matter [e.g., 
eqs.~(\ref{eq:CON1}) and (\ref{eq:CON2})]. For instance, in the special case of 
an ideal gas, one has $\Gamma_2 = c_P/c_V$, where $c_V$ is the specific heat
at constant volume (see \S9.14 in \cite{cg68} for further details). 

It should be noted that, although we derived eqs.~(\ref{eq:LEDOUX}) and 
(\ref{eq:SCHWARZ}) taking $\beta = 0$ (i.e., in the adiabatic case), it is also 
possible to arrive at these results on the basis of the dispersion relation 
eq.~(\ref{eq:DISPER}): as shown in \cite{sc05}, solutions of this equation 
containing a positive real part (i.e., indicating a growing perturbation and 
hence the onset of instability) are obtained if either eq.~(\ref{eq:LEDOUX})  
or eq.~(\ref{eq:SCHWARZ}) are satisfied, and also if 

\begin{equation} 
\frac{d\ln\mu}{dr} > 0.   
\label{eq:RAYLEIGH}
\end{equation}

\noindent This is the well-known {\em Rayleigh-Taylor instability}, which 
basically states that layers of matter with a higher molecular weight cannot 
exist in equilibrium on top of layers with lower molecular weight. In the 
case of low-mass stars, it has recently been suggested (\cite{peea06}), on
the basis of 3D models, that 
this instability may be important for reliable predictions of the stellar 
nucleosynthesis and Galactic evolution of $^3{\rm He}$, which plays a key 
role in the context of empirical tests of Big Bang nucleosynthesis 
(e.g., \cite{nhea95}, \cite{koea95}). On the other hand, it should be
noted that \cite{peea06} did not actually follow the mixing through the 
radiative zone above the hydrogen shell, but only speculated that such 
mixing should reach the convective envelope on a short timescale. Moreover,
they started their calculations with a model already well up the RGB in 
which previous mixing was not considered (because these models were 
carried out in 1D). Clearly, additional work will be required before 
putting their results on a firmer footing.

\vskip 1.5cm

\noindent{\bf The Temperature Gradient under Convective Stability} 

The Schwarzschild and Ledoux criteria tell us, for each point in the star, 
whether it is in radiative equilibrium or not. If it is, the temperature 
gradient is immediately given by eq.~(\ref{eq:RADGR}). What is the temperature 
gradient when radiative equilibrium does {\em not} hold~-- i.e., when convection 
also becomes an efficient mechanism of energy transport? 

In order to answer this question, we must study in more detail the properties 
of the convective bubbles shown in Fig.~\ref{fig:bubble}. More specifically, 
we must know the excess heat contained in the bubble, as well as its 
displacement speed and, of course, the distance traversed 
before it imparts its excess heat to its surroundings. The way this is 
usually accomplished, in stellar interiors work, is using the so-called 
{\em mixing length theory} of convection, as originally 
formulated by \cite{ev53,ebv58} (for alternative formulations of this 
theory, see also, e.g., \cite{lhea65,bp69}).

Turbulent convection is one of the most complicated problems in physics. 
However, 
within the scope of the mixing length theory, a number of approximations 
are made which make the problem more tractable. In particular, one first 
assumes that the full spectrum of eddies that normally characterizes a 
turbulent flow can be described in terms of a single, ``representative'' 
bubble. Second, one assumes that this bubble will traverse a distance 
$\ell$~-- the mixing length~-- before finally dissolving and giving out 
its excess heat to the surroundings. Third, it is assumed that the bubble 
preserves its physical identity throughout its displacement by the distance 
$\ell$. Fourth, the bubble is also assumed to keep pressure equilibrium 
with its surroundings during the whole process. Fifth, viscous forces are 
assumed to be negligible. These and other approximations are discussed, 
for instance, in \cite{lw58,cg68,es71,jc76,vc00a,chea04}.  

Let us first compute the excess heat $\Delta Q$ contained, per unit volume, 
within this convective element or ``bubble'' (Fig.~\ref{fig:bubble}). Since the 
bubble keeps pressure equilibrium with its surroundings until the moment 
when it gives out its excess heat, we have 

\begin{equation} 
 \Delta Q = \rho c_P \Delta T,   
\label{eq:DQ}
\end{equation}

\noindent where $\Delta T$ is $T_{\rm bub}-T_{\rm sur}$. Note that $\Delta T$ 
can also be written as 

\begin{equation} 
\Delta T  =  \left[\left(\frac{dT}{dr}\right)_{\rm bub} - \left(\frac{dT}{dr}\right)_{\rm sur}\right]\Delta r; \label{eq:BUB-SUR0}  
\end{equation}

\noindent therefore, defining 

\begin{equation} 
  \Delta\nabla T \equiv  \left(\frac{dT}{dr}\right)_{\rm bub} - \left(\frac{dT}{dr}\right)_{\rm sur}, 
\label{eq:BUB-SUR4}
\end{equation}

\noindent one finds

\begin{equation} 
\Delta T  =  \Delta\nabla T\Delta r, \label{eq:BUB-SUR1}
\end{equation}

\noindent and therefore

\begin{equation} 
 \Delta Q = \rho c_P \Delta\nabla T\Delta r.    
\label{eq:DQ2}
\end{equation}

We note, in passing, that $\Delta\nabla T$ can also be written as

\begin{eqnarray} 
\Delta\nabla T  & = & \left[\left(\frac{dT}{dP}\right)_{\rm bub} - \left(\frac{dT}{dP}\right)_{\rm sur}\right]\frac{dP}{dr} \\
         & = & \left[\left(\frac{d\ln T}{d\ln P}\right)_{\rm bub} - \left(\frac{d\ln T}{d\ln P}\right)_{\rm sur}\right]\frac{T}{P}\frac{dP}{dr} 
                                                                                              \label{eq:BUB-SUR0b} \\
         & = & \left(\nabla_{\rm bub}-\nabla_{\rm sur}\right)\frac{T}{P}\frac{dP}{dr} \label{eq:BUB-SUR} \\
         & = & \left(\nabla_{\rm sur}-\nabla_{\rm bub}\right)T\frac{\Delta r}{\lambda_P}. \label{eq:BUB-SUR5} 
\end{eqnarray}

Before proceeding, we must emphasize that the temperature gradients that appear 
in these equations are (in general) {\em not} the same as those appearing in 
eq.~(\ref{eq:DMT}). The four gradients should be interpreted as follows:  
$\nabla_{\rm rad}$ is the gradient that would be present under radiative 
equilibrium. $\nabla_{\rm ad}$ is the temperature gradient in the bubble if 
its displacement were adiabatic, i.e., if it did not exchange any heat with 
its surroundings during its ascent. $\nabla_{\rm bub}$ is the actual temperature 
gradient in the bubble, which will be equal to $\nabla_{\rm ad}$ only if the 
bubble's displacement is adiabatic. Finally, $\nabla_{\rm sur}$ is the 
temperature gradient in the surroundings of the bubble~-- in the preceding 
equilibrium discussion, this meant the radiative gradient, since we were 
discussing under which conditions the bubble's initial (small) motion would 
{\em lead} to a convective instability; in the present discussion, on the 
other hand, this stands for the {\em actual} temperature gradient that is 
present in the surroundings of the convective element, when the convective 
instability is {\em already} fully developed.   

Now the actual heat {\em flux} depends on the velocity of the bubble, which 
can in principle be computed by integrating the equation of motion 
[i.e., eq.~(\ref{eq:ARCHIM-FORCE-0}) or 
(\ref{eq:ARCHIM-FORCE})]. Instead of doing this, we note that the average 
force acting on the bubble (per unit volume), while it traverses the distance 
$\Delta r$, is given by 

\begin{equation} 
  \langle \mathcal{F} \rangle \simeq \frac{1}{2} g \Delta\rho,
\label{eq:AV-FORCE}
\end{equation}

\noindent where the factor 1/2 is due to the fact that $\Delta\rho/2$ is the 
average density difference between the bubble and the surroundings, since we 
assume that, initially, the bubble and the surroundings are basically 
indistinguishable (see Fig.~\ref{fig:bubble}) and, moreover, 
eq.~(\ref{eq:ARCHIM-FORCE}) implies that $\Delta\rho$, to first order, 
increases linearly with $\Delta r$. 
Now, in analogy with eq.~(\ref{eq:BUB-SUR0b}), 
$\Delta\rho$ can be computed from 

\begin{equation} 
\Delta\rho = 
\left[\left(\frac{d\ln\rho}{d\ln P}\right)_{\rm bub} - \left(\frac{d\ln\rho}{d\ln P}\right)_{\rm sur}\right]\frac{\rho}{P}\frac{dP}{dr}\Delta r.
\label{eq:BUB-SUR2}
\end{equation}

The convective energy flux can be computed from the foregoing equations more 
generally, but it is instructive to see what happens when we assume that the 
motion of the bubble {\em is} adiabatic, and ignoring molecular weight 
gradients (which, in any case, convection quickly tends to smooth out). 
In this case, keeping in mind eq.~(\ref{eq:AD-GAMMA}) and recalling that 
$P\propto \rho^{\Gamma_2}$ for an adiabatic transformation, one finds  

\begin{equation}
  \left(\frac{d\ln\rho}{d\ln P}\right)_{\rm bub} = \left(\frac{d\ln\rho}{d\ln P}\right)_{\rm ad} = 
  \frac{1}{\Gamma_2} = 1 - \nabla_{\rm ad}.
  \label{eq:DLNRHOAD}
\end{equation}
  
\noindent Evaluating this derivative for the surroundings is more complicated, 
since here we have to resort to the more general relation eq.~(\ref{eq:DMEOS0}).
In this case, one finds 

\begin{equation}
  \left(\frac{d\ln\rho}{d\ln P}\right)_{\rm sur} = \frac{1}{\chi_{\rho}} -
    \frac{\chi_{T}}{\chi_{\rho}} \left(\frac{d\ln T}{d\ln P}\right)_{\rm sur} - 
    \frac{\chi_{\mu}}{\chi_{\rho}} \left(\frac{d\ln\mu}{d\ln P}\right)_{\rm sur}.
  \label{eq:DLNRHOSUR}
\end{equation}

\noindent Therefore, in this case, ignoring molecular weight gradients, and 
recalling the definition eq.~(\ref{eq:BAS5}), one finds

\begin{equation}
  \left(\frac{d\ln\rho}{d\ln P}\right)_{\rm sur} = \frac{1}{\chi_{\rho}} -
     \frac{\chi_{T}}{\chi_{\rho}} \nabla_{\rm sur}. 
  \label{eq:DLNRHOSUR2}
\end{equation}

Dropping subscripts for the surrounding quantities, one then has 

\begin{eqnarray} 
\Delta\rho & = &
\left[\left(1 - \nabla_{\rm ad}\right) - 
\left(\chi_{\rho}^{-1}-\frac{\chi_{T}}{\chi_{\rho}} \nabla\right)\right]\frac{\rho}{P}\frac{dP}{dr}\Delta r \\ 
& = &
\left[\left(1-\chi_{\rho}^{-1}\right)+\left(\frac{\chi_{T}}{\chi_{\rho}}\nabla-\nabla_{\rm ad}\right)\right]\frac{\rho}{P}\frac{dP}{dr}\Delta r \\
 & = & 
-\left[\left(1-\chi_{\rho}^{-1}\right)+\left(\frac{\chi_{T}}{\chi_{\rho}}\nabla-\nabla_{\rm ad}\right)\right]\rho\frac{\Delta r}{\lambda_{P}}. 
\label{eq:BUB-SUR3}
\end{eqnarray}

This takes on a particularly simple form in the case of an ideal gas, when 
${\chi_{T}} = {\chi_{\rho}} = 1$. In this case, eq.~(\ref{eq:BUB-SUR3}) reduces to
(see eq.~\ref{eq:BUB-SUR5}, and recall that we are assuming that the bubble's 
displacement is adiabatic)

\begin{equation}
\Delta\rho  =  - \left(\nabla-\nabla_{\rm ad}\right) \rho\frac{\Delta r}{\lambda_{P}} 
            =  - \frac{\rho}{P}\Delta\nabla T \Delta r, 
\label{DRHOID}
\end{equation}

\noindent so that the average force (eq.~\ref{eq:AV-FORCE}) becomes

\begin{equation} 
 \langle \mathcal{F} \rangle \simeq \frac{g \rho}{2T} \Delta\nabla T\Delta r.
\label{eq:AV-FORCE2}
\end{equation}

\noindent At a given instant of time, a typical bubble will have traversed a
distance given by $\Delta r \approx \ell/2$, so that the average velocity 
can be computed from  

\begin{equation} 
 \langle \mathcal{F} \rangle \Delta r = \frac{1}{2}\rho\langle v^2\rangle
   \,\,\,\,\, \Rightarrow \,\,\,\,\,
 \langle v^2 \rangle \simeq \frac{g}{T} \Delta\nabla T \frac{\ell^2}{4}.
\label{eq:AV-VEL}
\end{equation}
 
\noindent Therefore, from eqs.~(\ref{eq:DQ2}) and (\ref{eq:AV-VEL}), and recognizing 
that there are about as many ``hot'' bubbles moving upwards as there are ``cold'' 
bubbles moving downwards, the convective energy flux is given by 

\begin{equation} 
 F_{\rm conv} = 2 v \Delta Q 
\simeq \rho c_P \frac{\ell^2}{4}\left(\frac{g}{T}\right)^{\frac{1}{2}}\left(\Delta\nabla T\right)^{\frac{3}{2}}.
\label{eq:CONV-FLUX}
\end{equation}

\noindent Except for a numerical factor of order unity~-- which is not of
primary relevance in the context of the mixing length theory, since the value 
of $\ell$ is still unspecified and must actually be calibrated by comparing 
the stellar model results (i.e., the predicted radii and effective 
temperatures) with the observations (i.e., the measured values for the 
Sun at its present age)~-- this expression is essentially identical to 
eq.~(21.14) in \cite{ts72}, or to eq.~(7.7) in \cite{ms58}, or also to 
eq.~(14.31a) in \cite{cg68}. 

In the convective cores of intermediate- and high-mass stars, where energy 
generation proceeds through the CNO cyle, the triple-$\alpha$ process, or 
even more advanced energy generation cycles (e.g., \cite{dc68,rr88}), 
one finds that the value of $\Delta\nabla T$ appearing in this equation 
is very small. This means that convection is very efficient in these cases, 
but a minor superadiabaticity (i.e., deviation from the adiabatic regime) 
being needed in order to transport the excess heat from the energy-generating 
regions of the stars. Under such circumstances, since the two gradients are 
so similar, it is sufficient to take the adiabatic value for the actual 
temperature gradient in eq.~\ref{eq:BAS4}. 
In the case of low-mass stars, this actually happens 
during the HB phase: for instance, adopting typical values for the core 
of an HB star from Fig.~11 in \cite{ii71}, one finds values of 
$\Delta\nabla T \approx 2\times 10^{-10}\,\,{\rm K\,cm}^{-1}$, compared 
with an actual temperature gradient some six orders of magnitude higher, 
namely, $dT/dr \approx 6\times 10^{-4}\,\,{\rm K\,cm}^{-1}$. 

However, in order to properly describe the temperature profiles 
of the convective {\em envelopes} of low-mass stars, eq.~(\ref{eq:CONV-FLUX}) 
must be solved, together with the radiative transfer equation, for the 
{\em actual} 
temperature gradient. In this case, the total flux is given by the combination 
of radiative and convective fluxes, thus: 

\begin{equation} 
 F = F_{\rm rad} + F_{\rm conv} = \frac{L}{4\pi r^2} = 
-\frac{4ac}{3{\overline{\kappa_{\rm R}}\rho} T^3}\frac{dT}{dr} +
\rho c_P \frac{\ell^2}{4}\left(\frac{g}{T}\right)^{\frac{1}{2}}
\left[\left(\frac{dT}{dr}\right)_{\rm ad}-\left(\frac{dT}{dr}\right)\right]^{\frac{3}{2}},
\label{eq:TOTAL-FLUX}
\end{equation}

\noindent where we have used the definition of $\Delta\nabla T$ as given  
in eq.~(\ref{eq:BUB-SUR4}), dropping the subscript 
``sur'' for the actual temperature gradient, and assuming that the bubble does 
move along an adiabat in parameter space. 

In order to solve this equation for the temperature gradient, one must know 
the mixing length $\ell$. What value does $\ell$ take on? In the mixing length 
theory, this is usually taken as proportional to the pressure scale height 
$\lambda_P$, so that 

\begin{equation}
  \ell \equiv \alpha_{\ell} \lambda_P, 
\label{eq:MLP}
\end{equation}

\noindent which corresponds to the definition of the {\em mixing length parameter} 
$\alpha_{\ell}$. The latter's value will depend on one's exact implementation of the 
mixing length theory, and even on the detailed opacity tables used [e.g., 
\cite{acea95}; see the dependence on ${\overline{\kappa_{\rm R}}}$ in 
eq.~(\ref{eq:TOTAL-FLUX})], so that it may be quite risky to compare values obtained 
by different authors without first making sure, among other things, that the 
numerical factor appearing in eq.~(\ref{eq:CONV-FLUX}) is the same for all, and 
that similar opacity tables have been used. 
In any case, typical values fall in the range $1.5 \lesssim \alpha_{\ell} \lesssim 2.5$, 
and are obtained by following the evolution of a one-solar-mass star with the 
same chemical composition as the Sun until the present (i.e., for $4.57\pm 0.05$~Gyr; 
see \cite{bp95,ms02}), and checking which value of $\alpha_{\ell}$ better matches the 
observed properties of the Sun~-- especially its radius and temperature, which 
depend crucially on the temperature profile of its outer envelope. 
One then {\em assumes} that 
the mixing length parameter is the same for every star, irrespective of its 
mass, chemical composition, or evolutionary phase. One may suspect that this may 
be too crude an approach for state-of-the-art, quantitative stellar evolutionary 
modelling~-- but in fact, this is still the procedure that is followed by the vast 
majority of workers in the field, and it does give, with virtually no additional
assumptions, a surprisingly good description of the observations for stars in 
globular clusters (e.g., \cite{fs99,rpea02,ffea06}).

\vskip 0.5cm

\noindent{\bf Overshooting, Semiconvection, and ``Breathing Pulses''} 

The criteria we have derived above for the placement of the boundaries of  
a convection zone (eqs.~\ref{eq:LEDOUX} and \ref{eq:SCHWARZ}) are strictly 
{\em local}, in the sense that they depend exclusively on the physical 
properties at the place we are studying. Nature, however, is more 
complicated than that; in particular, convective bubbles that are approaching 
the convective boundary set by either the Ledoux or the Schwarzschild criteria 
may retain sufficient momentum that they may be able to ``overshoot'' into 
the region that, according to those criteria, should be in radiative equilibrium. 
Since the bubble's velocity at the formal convective border, and hence its 
ultimate fate, clearly depends on what happened to it {\em before} it reached 
this point, a description of 
overshooting must involve {\em non-local} phenomena. 
In this sense, convective overshooting
is most frequently treated using the formalism developed by Roxburgh 
(\cite{ir65,ir78,ir89}). In its simplest form (\cite{ir78}), 
the actual extent of a convective region 
is determined by the values of $r_{\rm inn}$ and $r_{\rm out}$ (inner and outer 
radii of the convective region, respectively) which satisfy the following 
equation (the so-called ``Roxburgh criterion''):  

\begin{equation}
  \int_{r_{\rm inn}}^{r_{\rm out}} \left(L_{\rm rad}-L\right)\frac{1}{T^2}\frac{dT}{dr}dr = 0,  
\label{eq:ROX1}
\end{equation}

\noindent where $L_{\rm rad}$ is the portion of the total luminosity that is 
carried away in the form of radiation. This relatively simple equation is 
valid only if viscous dissipation can be neglected. The situation becomes much 
more complex if viscous dissipation is non-negligible; unfortunately, all  
that is known about the viscous dissipation tensor (eq.~6 in \cite{ir89}) 
is that it has a lower bound of zero, which brings about large uncertainties 
in models with convective cores. 

In the case of intermediate- and high-mass 
main-sequence stars, which do have convective cores, overshooting may lead 
to a substantial increase in the size of the convective core compared to that 
predicted by the Ledoux and Schwarzschild criteria, thereby also affecting 
their predicted lifetimes and evolutionary paths. One must accordingly 
calibrate the convective overshooting ``efficiency'' when computing 
evolutionary models for stars with convective cores. 
A commonly adopted procedure is to parameterize 
the convective overshoot in terms of the local pressure scale height $\lambda_P$ 
(see, e.g., \cite{ccea92} and references therein). Another, more sophisticated
approach has recently been suggested by \cite{dvea06}, who provide  
the following equation that must be solved for the actual extent of the 
convective core radius $r_{\rm cc}$: 

\begin{equation}
  \int_0^{r_0} f_{\rm over}\left(L_{\rm rad}-L\right)\frac{1}{T^2}\frac{dT}{dr}dr +
  \int_{r_0}^{r_{\rm cc}} \left(2-f_{\rm over}\right)\left(L_{\rm rad}-L\right)\frac{1}{T^2}\frac{dT}{dr}dr = 0,  
\label{eq:ROX2}
\end{equation}

\noindent where $r_0$ is the ``classical'' core boundary (i.e., as given 
by eq.~\ref{eq:LEDOUX} or \ref{eq:SCHWARZ}), and $f_{\rm over}$ is a free 
parameter that is evaluated by comparison with the observations, and which 
is, in general, a function of mass and chemical composition (see \cite{dvea06} 
for additional details).   

As pointed out by \cite{sd72}, theoretical investigations show that the 
core-helium burning phase of low-mass stars may be characterized initially 
by convective core overshooting, and later by the formation of a 
{\em semiconvective zone} surrounding the convective core. What happens in 
this phase is that the opacity of the matter in the convective core becomes 
higher and higher as He is transformed into C, as a consequence of the 
increasing contribution of the free-free process with increasing carbon 
abundance (\cite{vcea71a}). 
Since a chemical composition discontinuity is progressively built up between 
the C-enriched core and the He-rich envelope, a discontinuity in 
${\overline{\kappa_{\rm R}}}$ also results. This, in turn, leads 
to a discontinuity in the radiative gradient $\nabla_{\rm rad}$ accross 
the boundary of the convective core (see eq.~\ref{eq:RADGR}). Convective 
overshooting is then assumed to take place, the convective core 
increasing in size until convective neutrality is restored at its edge. 
However, once the convective core exceeds a certain size, the radiative 
gradient reaches a minimum and then increases with increasing distance 
from the center. According to canonical theory, the end 
result is the development of a {\em partially mixed}, convectively neutral 
(i.e., with $\nabla_{\rm rad} = \nabla_{\rm ad}$), ``semiconvective'' region 
outside the ``canonical'' convective core. Exceedingly clear descriptions 
of the entire process have been provided in \cite{as90,as94b}. 

The semiconvective phenomenon has been known for almost 50 years, having 
first been identified by Schwarzschild \& H\"arm in their study of the 
evolution of high-mass stars (\cite{sh58}). In it, they described the 
occurrence of a zone~-- the semiconvective zone~-- just outside the 
convective core ``in which the convection 
is so slow that it does not contribute to the energy transport but is fast 
enough to modify the composition, so that convective neutrality is maintained 
in every layer of this zone.'' Naturally, since they were dealing with the 
convective cores of high-mass MS stars, in their work they only tackled 
the phenomenon of {\em hydrogen} semiconvection. In low-mass stars, on
the other hand, semiconvection takes place in the core He-burning phase, 
and may accordingly be termed helium semiconvection. Interestingly, 
in the former case semiconvection arises because the opacity of the 
hydrogen-rich matter just outside the convective core is higher than 
just inside the convective core, whereas the opposite is true for 
the convective overshooting that leads to semiconvection in HB stars. 
For yet another example of the occurrence of semiconvection in low-mass
stars, the reader is referred to the studies of \cite{ir82a,ir82b,hi89} 
on the role played by semiconvection on the AGB phase. 

By bringing He-rich material into the core, a semiconvective zone 
enables a low-mass star to burn helium that is present over a much 
wider range in mass than would otherwise have been possible, thereby 
leading to a significant increase in the HB lifetime and also dramatically 
impacting the morphology of HB evolutionary tracks in the HRD 
(e.g., \cite{vcea71a,vcea71b,dm72,rf72,sd72,sg74,sg76,rfp88,as94b}). 
From the point of view of a fluid dynamics theoretician, on the other 
hand, ``the 
still unsolved problem''~-- in the words of Canuto (\cite{vc00b})~-- ``is 
as follows: what are the values of $\nabla$ and $\nabla_{\mu}$ 
in such a zone?'' In addition, canonical 
theory simply assumes that the composition distribution within the 
semiconvective zone instantaneously adjusts to insure convective 
neutrality, without any attempt to follow the time evolution of the 
convective layer. The reader is referred 
to \cite{vc99} for a recent, detailed discussion of this problem, and to 
\cite{wm95} for 2D hydrodynamical calculations and comparison of the results 
with commonly adopted prescriptions for the treatment of semiconvection in 
stellar interiors (see also \cite{gb06}). 

An important problem related to the semiconvection mechanism operating in 
HB stars are the so-called ``breathing pulses,'' which lead to abrupt 
{\em enhancements} in the core helium abundance as a consequence of sudden 
outward movements of the edge of the convective core when He is approaching 
exhaustion in the core (i.e., when $Y_{\rm core} \lesssim 0.1$; 
e.g., \cite{sd72,sd73,vcea85,cm93,as94b}). 
The main question here is whether these ``breathing pulses'' are real or 
instead an artifact of the canonical assumptions used to treat convection 
and mixing in stellar interiors (e.g., \cite{as90,dr93,as94b}). 
Since these pulses significantly 
increase the amount of He fuel that is burnt on the HB phase, they 
accordingly decrease the amount of He that is available for burning on 
the AGB phase; therefore, they may crucially affect the predicted number 
ratios between AGB and HB stars (e.g., \cite{rfp88}). 
While inclusion of semiconvection is indeed crucial for correctly 
predicting the observed AGB-to-HB number ratio 
(e.g., \cite{rbea85,rfp88,fcea89}), comparisons between models 
that allow for the breathing pulses and the observations strongly suggest 
that these pulses do not occur in real stars (e.g., \cite{scea01}). 
However, if time-dependent overshooting is adopted as opposed to canonical 
semiconvection, breathing pulses arise naturally and the corresponding 
models are still consistent 
with the observations (\cite{as94b}). Indeed, in the HB computations by 
\cite{as94b}, in which time-dependent overshooting was considered, the overall 
results (HB lifetime, evolutionary track morphology) were found to be in 
excellent agreement with those using canonical semiconvection but suppressing 
breathing pulses, with one important difference: the breathing pulses, 
by perturbing the interior structure as they redistribute carbon from 
the central He-burning regions, cause fluctuations in the stellar radius, 
and thus in the predicted RR Lyrae pulsation periods, which may be related 
with the~-- often erratic~-- period changes observed in RR Lyrae stars 
(\cite{sr79,as94b}; see also \cite{ac98} and \cite{mpea02}, and \S9 in 
\cite{mc05} for a review of the available observations). 
On the other hand, \cite{abea86} have argued that convective breathing 
pulses, and in fact even semiconvection itself, are an artifact of the 
use of a local theory of convection, disappearing when a non-local 
formalism is properly adopted. However, the model of convection upon 
which the reasoning by \cite{abea86} is based has been strongly 
questioned (\cite{ar87}). The reader is referred to \cite{ldea96} and 
\cite{cc99} for recent discussions on this hotly debated topic. 

To close, we note that semiconvection at the base of the convective 
{\em envelope} has recently been found in evolved solar models (at an 
evolutionary age of 6.5~Gyr) by \cite{jbea01}, as a consequence of diffusion 
increasing the abundances of heavier elements (hence the opacity) below the 
base of the convective zone. (See \S6.2.3 in the monograph by \cite{ms02} 
for a review of overshooting at the base of the solar convective envelope.)
To the best of the present author's knowledge, a similar phenomenon has not 
yet been described/included in other studies of low-mass stellar evolution.

\subsubsection{Conductive Transport}
\label{sec:cond}

Besides radiation and convection, energy can also be transported from one 
place in the star to another by the movement of individual particles 
(i.e., electrons, protons, neutrons, $\alpha$-particles, etc.). In 
practice, given their low masses (hence high velocities) and the high 
degree of ionization in stellar interiors, only electrons play a truly 
significant role in this regard. 

Still, under normal conditions, it can be shown that energy transport by 
conduction is much smaller than by either radiation or/and convection, 
so that it is safe to neglect its contribution when computing stellar 
structure models. This rule of thumb, however, breaks down under one 
circumstance which is of crucial importance in the life of a low-mass 
star, namely, the development of {\em electron degeneracy}. When matter 
becomes degenerate~-- which is what happens in the cores of RGB and AGB 
stars and in white dwarfs~-- the low-lying energy states all become filled, 
which effectively prevents electrons from undergoing frequent collisions 
with other ions or electrons (since if they did collide their energies 
might change, which cannot frequently happen since only high-lying 
energy states have not yet been filled). As a consequence, the mean 
free path of an electron increases dramatically, and a typical 
electron may be able to travel a large distance before finally 
colliding and releasing its excess energy. A strongly degenerate 
gas is often approximated by a zero-temperature Fermi gas; however, 
this is not a sufficiently good approximation in the case of low-mass 
RGB and AGB stars, in which degeneracy is not so strong. Note that, 
in the evolutionary phases preceding the RGB, the gas in the deep 
stellar interior can be approximated as a fully ionized perfect gas. 

Conductive opacity $\kappa_{\rm c}$ 
is defined in such a way that the energy flux by 
electron conduction takes on a form similar to that which applies to 
radiative transport (eq.~\ref{eq:FLUXT}), namely, 

\begin{equation} 
F_{\rm c} = -\frac{4ac}{3\kappa_{\rm c}\rho} T^3 \frac{\partial T}{\partial r}, 
\label{eq:FLUXCOND}
\end{equation}

\noindent where $F_{\rm c}$ is the conductive energy flux. Therefore, in 
a region where conduction is important, the total energy flux is given by 

\begin{equation} 
F = F_{\rm rad} + F_{\rm c} = -\frac{4ac}{3\kappa_{\rm equiv}\rho} T^3 \frac{\partial T}{\partial r}, 
\label{eq:TOTALFLUX}
\end{equation}

\noindent where the ``equivalent opacity'' $\kappa_{\rm equiv}$ is given 
by 

\begin{equation} 
\frac{1}{\kappa_{\rm equiv}} = \frac{1}{\overline{\kappa_{\rm R}}} + \frac{1}{\kappa_{\rm c}}.
\label{eq:OPEQUIV}
\end{equation}

\noindent As can be seen, the radiative and conductive opacities add as do 
the resistances of two resistors associated in parallel in a simple electric 
circuit; this means that energy flow proceeds primarily through the ``easier'' 
of the two channels (i.e., the one with the lower opacity), just 
like electricity will flow primarily through the resistor with the lower 
resistance. Note that this rule is not valid for the computation of either 
the radiative or conductive opacities individually; indeed, the former is 
the result of the summation over many different physical processes, such as 
bound-bound (or line) transitions, bound-free transitions (photoionization), 
free-free transitions (inverse bremsstrahlung), molecular transitions, 
Thomson scattering, etc. 
(see, e.g., eq.~8.18 in \cite{dg05}), whereas the latter, according to the 
so-called {\em Matthiessen rule}, is also given by the sum of electron-ion 
and electron-electron contributions (see, e.g., \cite{jz60}) -- though it 
should be noted that this rule is, in fact, an approximation which is 
strictly valid only under conditions of strong electron degeneracy (see 
\cite{jz60,hl69}). 

Over the past several decades, the most frequently adopted recipes for conductive 
opacities in stellar interior calculations have been those by Hubbard \& 
Lampe (\cite{hl69}) and by Itoh and co-workers (e.g., \cite{niea83}). 
However, and as noted by \cite{mcea96}, these recipes are not strictly 
valid in the interiors of low-mass RGB stars, where matter is in an 
intermediate form between a gas and a liquid (Coulomb coupling parameter 
$\Gamma \lesssim 0.8$). The more recent calculations by \cite{ap99,apea99} 
have likewise not been strictly extended to these intermediate-coupling regimes, 
though an interpolation scheme was adopted which guarantees a smooth fitting 
accross the crucial region (see Fig.~17 in \cite{mc05}) between 
$\Gamma \rightarrow 0$ and $\Gamma > 1$. 

More recently, \cite{mc05} also noted that the available conductive 
opacity calculations have not taken into due account the fact that the 
matter in RGB interiors is {\em not} strongly degenerate, being better 
characterized instead by intermediate degeneracy levels (see Fig.~18 
in \cite{mc05}). This statement is especially applicable to the 
electron-electron component, which is of great importance for low-mass 
RGB interiors in particular (e.g., \cite{mcea96,mc05}). 

In order to provide conductive opacity calculations that should be more 
safely applicable to the conditions characterizing the interiors of low-mass 
RGB stars, \cite{scea07} have recently undertaken the task of revisiting the 
work by \cite{ap99,apea99}, addressing the limitations that were reviewed 
in the previous paragraphs and taking into account additional physical 
processes (\cite{sy06}) which had not been previously incorporated into 
such calculations. As a result, \cite{scea07} came up with revised 
conductive opacities which differ in some important  
respects from those provided in 
previous work for the physical conditions prevailing in the 
interiors of low-mass RGB stars (see Fig.~4 in their paper), and which 
present a non-negligible impact on such observables as the luminosity of 
the RGB tip, the luminosities of HB and RR Lyrae stars, and the predicted 
RR Lyrae pulsation periods (see \cite{scea07} for further details). These
revised conductive opacities are also available on the web.\footnote{
{\tt http://www.ioffe.ru/astro/conduct/index.html}\label{foo:alex}}

\subsubsection{Some Additional Open Problems}
\label{sec:unc}

\noindent {\em Radiative opacities:} 
As already stated (Sec.~\ref{sec:rad}), great effort has been dedicated, 
over the past 15 years or so, to compute realistic tabulations of Rosseland 
mean opacities (and the accompanying EOS) for use in astrophysical applications. 
These efforts have been mainly led by the OPAL team at the Lawrence Livermore 
National Laboratory (e.g., \cite{ir96}; see also footnote~\ref{foo:opal}) and 
by the OP team, led by M. Seaton at the University College London 
(\cite{ms05,nbea05} and reference therein; see footnote~\ref{foo:op}). 
The opacities provided by these two groups have proven of great importance 
in explaining (and even predicting, as in the case of the EC~14026 non-radial 
pulsators) a variety of stellar pulsation-related phenomena (e.g., 
\cite{ac91,sk92,ns95,scea96,bvea97,amea07} and references therein). 
In the low-temperature regime, where molecular effects become important, 
the Alexander-Ferguson opacities (\cite{af94}) are generally to be preferred, 
but the recent updates and extensions by \cite{jfea05} should now be implemented 
as well.\footnote{{\tt http://webs.wichita.edu/physics/opacity/}\label{foo:lowt}} 
Account should be taken of possible deviations from scaled-solar 
abundances, since these may have dramatic impacts on the resulting stellar 
models (e.g., \cite{awea07}). 

The improved calculations provided by these groups have proved of key 
importance for the solution of several long-standing problems in stellar 
astrophysics. Yet, some significant differences between OPAL and OP results 
still exist, and may reach up to 25\% in some key regions of parameter 
space (\cite{sb04,amea07}), such as the driving regions in some types of 
pulsating stars (e.g., \cite{js06,amea07}), leading to significantly 
different results depending on which set of opacities is used. Some of 
these differences appear to be due to the different prescriptions for 
the EOS used in the different projects (\cite{sb04}). Needless to say, 
a resolution of the discrepancies that are present both at the radiative 
opacities and EOS fronts would be crucial to put the models of the structure, 
evolution, and pulsation of low-mass stars on a firmer footing. In this sense, 
attention should be called to the recent, provocative work by \cite{vtea96,vtea01} 
(and references therein), according to whom improvements in the treatment of
several important physical ingredients are still in order.

\vskip 0.5cm

\noindent {\em Convection beyond the Mixing Length Theory:} 
As can easily be realized from Sec.~\ref{sec:conv}, the currently most frequently  
adopted prescriptions for convection in stellar interiors are rather crude, 
this remaining, in fact, one of the most important sources of uncertainty 
in the computation of stellar models. In recent years, some attempts have 
been made to provide a more realistic description of convection for use in 
stellar interior calculations, going beyond the extremely simplified (and 
perhaps simplistic) mixing length formalism. In particular, \cite{cm91,cm92} 
have developed the so-called {\em full spectrum of turbulence} 
model which, unlike the mixing length theory, takes into account the 
existence of turbulent eddies of all sizes. In this theory, the concept 
of mixing length is also redefined; in particular, the authors advocate 
the use of $\ell \propto z$ in the computation of stellar models, where 
$z$ is the distance to the top of the convection zone as determined by the 
Schwarzschild criterion (eq.~\ref{eq:SCHWARZ}). 

More recently, 
\cite{vcea96} presented an extension of the original full spectrum of 
turbulence model, in which the processes that generate the turbulence were 
more carefully considered. Unfortunately, a free parameter $\alpha^{*}$ 
(namely, a constant of proportionality, presumably smaller than 0.2 in 
absolute value, between the mixing length and the pressure scale height 
$\lambda_P$ at the upper boundary of the convection zone; see eq.~[96] in 
\cite{vcea96}) still has to be adjusted to the data for the Sun in order 
to ``calibrate'' this model. 

In reality, both the mixing length and the full spectrum of turbulence 
models face a strong challenge when confronted with both hydrodynamical 
simulations and the available asteroseismological data for the Sun 
(see \S2.1.3 in \cite{cb00} for a recent review and additional 
references). A recent attempt at calibrating the 
free parameters of both the mixing length and full spectrum of turbulence 
theories on the basis of 2D hydrodynamical simulations has been provided by 
\cite{hglea99} (see also \cite{sa00}). 
According to their results, the parameter $\alpha_{\ell}$ 
in the former theory presents a variation from $\sim 1.3$ for F-type 
dwarfs to $\sim 1.75$ for K-type subgiants, there existing a plateau 
in the neighborhood of the Sun where $\alpha_{\ell}$ remains nearly 
constant. It remains to be seen whether these variations in $\alpha_{\ell}$
can be reconciled with the observations of globular cluster stars (e.g., 
\cite{rpea02,ffea06}). Also for the full spectrum of turbulence model, 
\cite{hglea99} find a significant variation in $\alpha^{*}$ depending 
on the spectral type, with absolute values ranging from slightly negative 
for the cooler stars up to 0.6 for K-type dwarfs (note that, according to 
\cite{vcea96}, such a value should not be larger than 0.2). 

In fact, and as also pointed out in 
\cite{cb00}, the mixing length formalism appears to provide a {\em better} 
description of the thermal profile of the solar convective region, as 
compared with the full spectrum of turbulence model. However, adding 
to an already confusing situation, \cite{vc00a} has challenged the 
interpretation of the results of the numerical simulations, pointing 
out important limitations of 2D (as compared to 3D) calculations and arguing
that such simulations in fact rule out the mixing length theory while not 
being inconsistent with the full spectrum of turbulence model. One way 
or another, it appears clear that the solution to this problem lies in 
the computation of realistic 3D hydrodynamical simulations, which may 
differ substantially from both 2D simulations and the mixing length or 
full spectrum of turbulence predictions (see \cite{ma07} for a very 
recent example).

\subsection{Thermonuclear Reaction Rates} 
\label{sec:nucreact}

Thermonuclear reaction rates are of paramount importance for realistic 
stellar model computations for two main reasons: first, it is the energy 
generated by these reactions that allows the star to shine during most of 
its lifetime; second, these reactions lead to dramatic changes in the 
chemical compositions of (especially) the innermost regions of the stars, 
which is the primary driver of changes over time in the stellar structure, 
and accordingly also the reason why stars go through different evolutionary 
phases in the course of their existence. 

In state-of-the-art evolutionary calculations, {\em nuclear reaction networks} 
are routinely employed which take into account many different nuclear reactions, 
thus allowing one to follow the time evolution of a variety of different 
nuclear species. In the case of low-mass stars, the most important reactions 
that must be included in such networks are obviously the H- and He-burning 
reactions, most important among these being the so-called {\em proton-proton 
(PP) chain} for main-sequence stars, the {\em CNO cycle} for RGB, HB, and 
AGB stars, and the {\em triple-$\alpha$ process} for the HB and AGB phases. 
In addition, the $^{12}{\rm C}(\alpha,\gamma)^{16}{\rm O}$ reaction becomes 
increasingly important as a He-burning mechanism towards the end of the HB 
phase, when He is approaching exhaustion and the triple-$\alpha$ process 
becomes less frequent (involving, as it does, the almost simultaneous 
combination of three $^4$He nuclei). Proton-capture nucleosynthesis may  
also be of considerable importance, particularly in view of observed 
abundance anomalies (especially among globular cluster stars) in such 
elements as Na and Al (\cite{rgea04} and references therein). 
Neutron-capture nucleosynthesis (the so-called 
slow neutron capture process, or simply $s$-process; see \cite{fhea56}
for the first usage of the term in the scientific literature, and 
\cite{ebea57,ac57} for more extensive developments of the idea) may also take 
place during the AGB phase of low-mass stars (e.g., \cite{ii82,osea06}), 
due to the strong neutron fluxes that may be provided by such reactions as 
$^{13}{\rm C}(\alpha,n)^{16}{\rm O}$ and 
$^{22}{\rm Ne}(\alpha,n)^{25}{\rm Mg}$,  
particularly in the course of thermal pulses (see \S\ref{sec:evolover}).

\begin{figure}[t]
  \includegraphics*[width=4.2in]{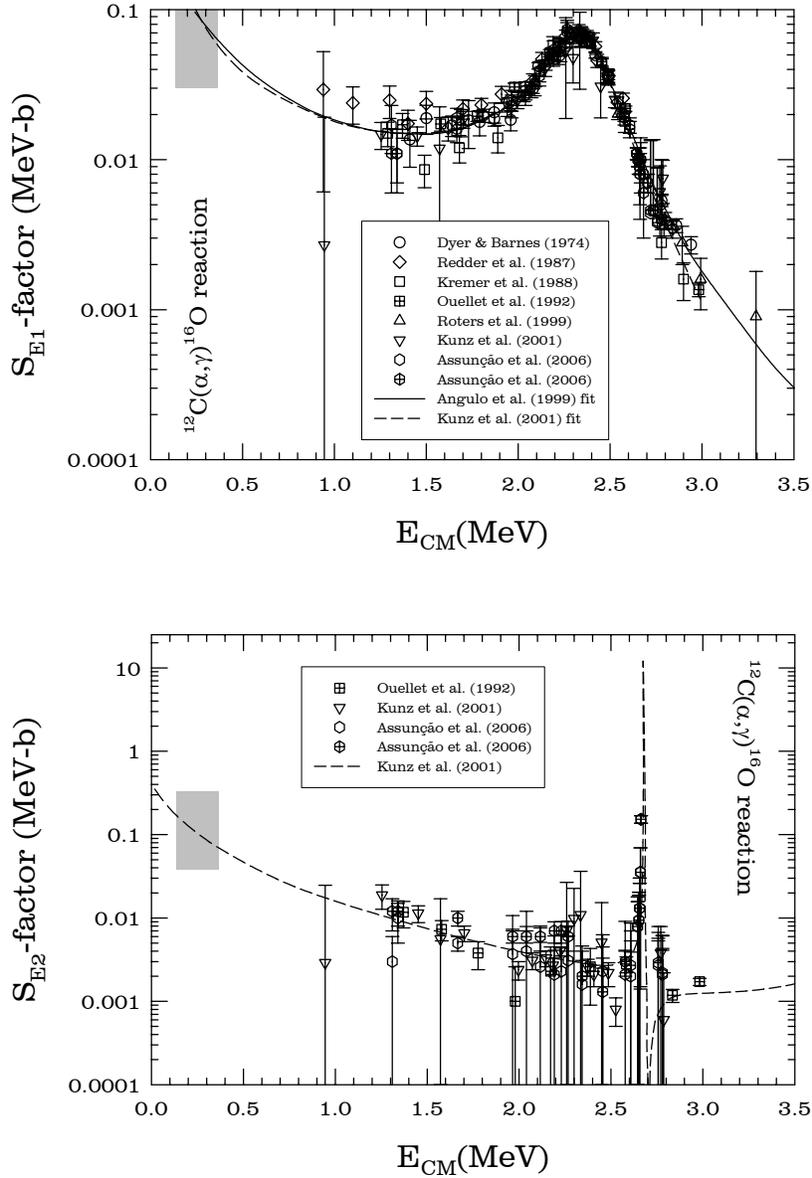}
  \caption{({\em top panel}) 
  Astrophysical $S$-factor (E1 component, in MeV-barn), for 
  the $^{12}{\rm C}(\alpha,\gamma)^{16}{\rm O}$ nuclear reaction as a
  function of the energy in the center-of-mass reference system (in 
  MeV). The sources of the indicated experimental data are shown in 
  the inset. The two 
  different symbols used in the Assun\c{c}\~{a}o et al. (2006) case 
  refer to two different solutions proposed by the authors (see their 
  paper for further details). 
  The fits to the data proposed in \cite{caea99} ({\em solid line})
  and in \cite{rkea01} ({\em dashed line}) are also shown. The region 
  of astrophysical interest, as indicated by the Gamow peak temperature 
  and $e$-width at a temperature $1.35\times 10^8$~K (a value typical 
  for He-burning at the end of the HB phase), is shown schematically 
  as a gray band. Note that the $y$-scale is logarithmic. 
  ({\em bottom panel}) 
  As in the top panel, but for the E2 component.
   }
      \label{fig:sfactor}
\end{figure}

All these reactions and cycles are reviewed in 
great detail in the monographs by Clayton (\cite{dc68}) and Rolfs \& Rodney 
(\cite{rr88}). Modern calculations most frequently adopt the nuclear reaction 
rates and $Q$ values that are provided in the NACRE 
database\footnote{{\tt http://pntpm.ulb.ac.be/Nacre/}\label{foo:nacre}} 
(\cite{caea99}), 
though on occasion computations are also provided which incorporate revised  
rates for some specific reactions that have been recommended in the 
nuclear physics literature (for recent examples, see, 
e.g., \cite{apea04,awea05,ac06}); conversely, it frequently happens that, on 
the basis of astronomical observations and astrophysical considerations, 
changes to the recommended values for some thermonuclear reaction rates are 
indicated (e.g., \cite{tm03,vd06} and references therein). In fact, one of 
the most spectacular astrophysical success stories ever came precisely 
from an argument of this type: Fred Hoyle and co-workers (\cite{fhea53}) 
realized that the observed (i.e., fairly high) amount of carbon in the 
Universe would be completely inconsistent with stellar nucleosynthesis
arguments, {\em unless} the $^{8}{\rm Be}(\alpha,\gamma)^{12}{\rm C}$ 
reaction proceeded through a (then unknown) {\em resonance} located very 
near the $^{8}{\rm Be} + \alpha$ threshold, i.e., at an energy around 
7.68~MeV. This reaction forms the second ``leg'' of the triple-$\alpha$ 
process, the first being 
$^4{\rm He} + ^4{\rm He}\,\rightleftarrows\, ^{8}{\rm Be}$
(\cite{eo51,es52}). 
The problem with this process, it was argued, is that $^{8}{\rm Be}$ very rapidly 
decays back into two $\alpha$ particles, thus making it exceedingly unlikely 
that substantial amounts of $^{12}{\rm C}$ can be formed without a resonance 
being present close to the $^{8}{\rm Be} + \alpha$ threshold. In addition to  
the corresponding $^{12}{\rm C}$ nuclear energy level proper, Hoyle et al. 
also predicted the precise quantum-mechanical properties (angular momentum 
and parity) of the corresponding $^{12}{\rm C}$ nuclear energy level. 
Amazingly, these purely astrophysical predictions were soon verified in 
laboratory experiments (\cite{ccea57}). Salpeter has recently published 
a very interesting review paper that describes these and other related 
developments in nuclear astrophysics from a historical perspective 
(\cite{es02}).   

The major uncertainty affecting the application of nuclear reaction rates 
that are measured in the laboratory to the realm of stellar interiors is 
the fact that, in the latter, those same reactions usually take place at 
much lower energies. This is a consequence of the fact that there are so  
very many particles available in a typical stellar interior that even if a 
minor fraction of them~-- i.e., those located towards the high-energy tail 
of the Maxwell-Boltzmann velocity distribution, or (more precisely) around 
the so-called ``Gamow peak''~-- are able to actually 
take part in nuclear reactions, still the required amount of energy to 
support the star can be generated in the process.  
The Gamow peak defines, by combining the competing 
effects of the high-energy tail of the Maxwell-Boltzmann distribution 
with the probability of tunneling through the corresponding Coulomb barrier, 
the most effective energy for thermonuclear reactions (see, e.g., \S4.3 in
\cite{dc68}, or \S4.2 in \cite{rr88}). In the laboratory, on 
the other hand, at similar energies, it is impractical to follow a 
sufficient number of events that might stand a chance of revealing even 
a bear minimum of such reaction events ever taking place. Therefore, in  
stellar astrophysics work, one is routinely forced to {\em extrapolate} 
laboratory measurements of the cross sections towards energies of relevance 
for stellar interiors (as indicated by the position of the Gamow peak), 
to rely on theoretical predictions of nuclear physics, and/or to use a 
combination of these two approaches. 
$^{12}{\rm C}(p,\gamma)^{13}{\rm N}$, the key reaction that starts the 
CNO cycle, is a classical example of a non-resonant 
reaction whose rate is primarily affected by this type of uncertainty 
(see, e.g., \S6.2.1 in \cite{rr88}).  

Of course, when carrying out such extrapolations, there is always the danger 
that previously unknown nuclear resonances may fall in the ``uncharted 
territory'' frequently represented by astrophysical energies~-- but this is 
a risk that must be assumed in stellar structure/evolution work. On the 
other hand, and as 
already pointed out, this is also one of the reasons why astrophysics may 
be used as an empirical probe of nuclear reactions, even beyond the domain 
within reach in a physicist's laboratory.

\subsubsection{The $^{12}{\rm C}(\alpha,\gamma)^{16}{\rm O}$ Reaction}
\label{sec:c12o16}

Probably the main source of uncertainty in the calculation 
of the structure and evolution of low-mass stars, as far as nuclear 
reactions rates are concerned, is the 
$^{12}{\rm C}(\alpha,\gamma)^{16}{\rm O}$ reaction~-- which, as 
already stated, becomes of increasing importance towards the end of the 
HB phase. At present, the cross section for this reaction, for typical 
He-burning temperatures, is 5-6 orders of magnitude below the experimental 
sensitivity (\cite{rkea02}). 

This reaction
proceeds through several different channels (see, e.g., Fig.~1 in 
\cite{rkea02}), the most important of which being called the E1 and 
E2 channels. In its essence, the E1 channel involves the low-energy  
tail of the $^{16}$O nuclear energy level that is located 2.418~MeV 
above the $^{12}{\rm C} + \alpha$ rest mass energy, plus a resonance 
that is located 45~keV {\em below} the latter (a so-called 
{\em subthreshold} resonance). The E2 channel, in 
turn, also involves a subthreshold resonance, namely, the one located 
245~keV below the $^{12}{\rm C} + \alpha$ rest mass energy, plus the 
direct-capture process into the $^{16}$O ground level. The fact that 
more than one resonance is at play in each case leads to 
{\em interference effects} between them. As shown in 
\cite{rr88} (see their Fig.~7.6), the resulting cross section 
depends on whether such interference effects are 
constructive or destructive. While for the E2 channel it is possible 
to determine the sign of the interference effects from measurements 
at relatively high energies (maximum interference is expected around 
3~MeV), in the case of the E1 channel measurements at energies 
around 1.4~MeV or lower are also required. From the available 
experimental data, it appears that constructive intereference is 
at play for both the E1 (see Fig.~7.10 in \cite{rr88}) and E2 cases 
(see Fig.~\ref{fig:sfactor}, and compare with Fig.~7.6, middle panel, 
in \cite{rr88}). 

To 
illustrate the present level of uncertainty in evaluating the reaction 
rate that should be used in stellar applications, in Fig.~\ref{fig:sfactor} 
({\em upper panel})
we plot, as an example, some of the available laboratory data for the E1 
branch, as compiled on the NACRE team web page, and supplemented by the 
more recent data from \cite{grea99,rkea01,maea06}. 
Here the so-called ``astrophysical $S$-factors'' are used as opposed to 
cross sections proper, the latter being obtained in terms of the former 
using the following expression:

\begin{equation} 
\sigma(E) = \frac{S(E)}{E} {\rm e}^{-2\pi\eta}, 
\label{eq:SFACTOR}
\end{equation}

\noindent where $E$ is the energy in the center-of-mass system, and 
$\eta$ is given by 

\begin{equation} 
\eta = \frac{Z_1 Z_2 e^2}{\hbar v},  
\label{eq:ETAS}
\end{equation}

\noindent where $Z_{1}$ and $Z_2$ represent the atomic numbers of nuclei 
1 and 2, respectively, $e$ is the electron charge, $\hslash = h/(2\pi)$ 
(with $h$ being the Planck constant), and $v$ is the relative velocity in 
the center-of-mass system. The goal here is to factor out the non-nuclear 
ingredients, such as the geometric de Broglie cross section 
$\pi\lambdabar^2 \propto 1/E$ and the probability of tunneling through 
the Coulomb barrier $\propto \exp(-2\pi\eta)$. Therefore, $S(E)$ is, by 
design, essentially the ``purely nuclear'' contribution to the cross 
section $\sigma(E)$. 

In Fig.~\ref{fig:sfactor} we also show, as a solid line, the 
(theoretically-motivated) fit to the data proposed by \cite{caea99}, 
as well as the similar fit more recently advanced by \cite{rkea01}. 
In addition, the range of energies of interest for low-mass stars, 
as given by the Gamow peak position and $e$-width [from eqs.~(4-47) and 
(4-52) in \cite{dc68}, respectively] at a temperature of 
$1.35\times 10^8$~K (as appropriate for the cores of helium-burning HB 
stars), is schematically shown as a gray band. While the fits proposed 
in \cite{caea99} and \cite{rkea01} are in good agreement in the region 
of interest, the situation is in fact more complex in the case of the E2 
component (see Fig.~\ref{fig:sfactor}, {\em bottom panel}), which leads 
to significant differences between the final recommended $S$-factors in 
\cite{caea99} and \cite{rkea01} (see Tables~1 and 4 and Fig.~3 
in \cite{rkea01}). In this sense, note, from Fig.~\ref{fig:sfactor},
that the E1 and E2 components have similar magnitudes in the energy range 
of interest for stellar interiors work. (All additional components, 
according to the bottom panel in Fig.~1 of \cite{rkea01}, are some three 
orders of magnitude smaller.)

\subsubsection{Screening Factors}
\label{sec:screen}

Before closing, we note that, since in the stellar interior the 
nuclei participating in the nuclear reactions are actually immersed in a 
plasma, the latter may actually become polarized, with electron ``clouds'' 
forming around the positive nuclei and thus leading to some ``shielding'' 
of the repulsive Coulomb potential~-- with the end result that the nuclear 
reactions proceed at faster rates than in the laboratory. This phenomenon 
is accounted for using the so-called {\em plasma screening factors} (see, e.g.,  
\S6.8 in \cite{dc68} for a simple treatment). In modern stellar structure 
work, screening factors are still most frequently taken from \cite{hdea73,hgea73}. 
However, and as previously pointed out in \cite{mcea96}, reportedly
more accurate prescriptions have been provided by \cite{ys89} (see their  
\S3). The impact of the Yakovlev-Shalybkov screening factors, to the best 
of the present author's knowledge, has unfortunately never been seriously 
evaluated in the context of the structure and evolution of low-mass stars.

\subsection{Boundary Conditions}
\label{sec:bound}

In order to integrate the set of eqs.~(\ref{eq:BAS1})-(\ref{eq:BAS4}), 
in addition to the adequate input physics indicated by 
eqs.~(\ref{eq:CON1})-(\ref{eq:CON5}), four suitable boundary conditions 
are obviously needed. Two of these~-- the {\em central boundary 
conditions}~-- are self-explanatory, namely (in Lagrangean notation): 

\begin{equation} 
r(m=0) = 0, \,\,\,\,\,\, L(m=0) = 0. 
\label{eq:CENTRAL}
\end{equation}

\noindent Though simple, these boundary conditions must be used 
with due care, or else they may lead to a singularity in the star center 
(see eqs.~\ref{eq:BAS1}-\ref{eq:BAS4}, and \S3.2 in \cite{sc05} for a recent
discussion). 

The other two boundary conditions come from the outer border 
of the star; basically, one needs to specify how pressure and temperature 
change with depth close to the stellar surface (defined as the point where 
the Lagrangean coordinate mass reaches its maximum value, corresponding 
to the total mass of the star). One cannot simply state the temperature 
and pressure at the ``actual'' surface of the star, since built-in into 
the aforementioned equations is the hypothesis of near isotropy of the 
radiation field, which breaks down more dramatically the closer one gets 
to the surface. For this reason, so-called {\em photospheric boundary 
conditions} must be used instead, which should ideally be based on 
detailed model atmospheres which take into account such potentially 
important effects as non-zero curvature, microturbulence, and deviations 
from local thermodynamic equilibrium. When such detailed models are 
lacking or/and practical considerations may prevent them from being 
implemented, the simpler Eddington boundary condition  (see \S226 in 
\cite{ae26}) is frequently employed. 

{\em Semi-empirical} boundary conditions are also frequently used. 
In this sense, a still very common approach is to use an analytical 
relationship between temperature and optical depth $\tau$ in the 
atmosphere, as given by Krishna-Swamy (\cite{ks66}), which reads 
as follows: 

\begin{equation}
T^4 = \frac{3}{4} T_{\rm eff}^4 \left(\tau + 1.39 - 0.815\,{\rm e}^{-2.54\,\tau}
      - 0.025\,{\rm e}^{-30\,\tau}\right), 
\label{eq:KRISHNA}
\end{equation}

\noindent where $T_{\rm eff}$ is the effective temperature. In practice, 
one assumes this equation to be valid up to a certain optical depth, at 
which point integration of the stellar interior equations must yield 
a temperature matching this result.

%
\begin{figure}[t]
  \includegraphics*[width=5in]{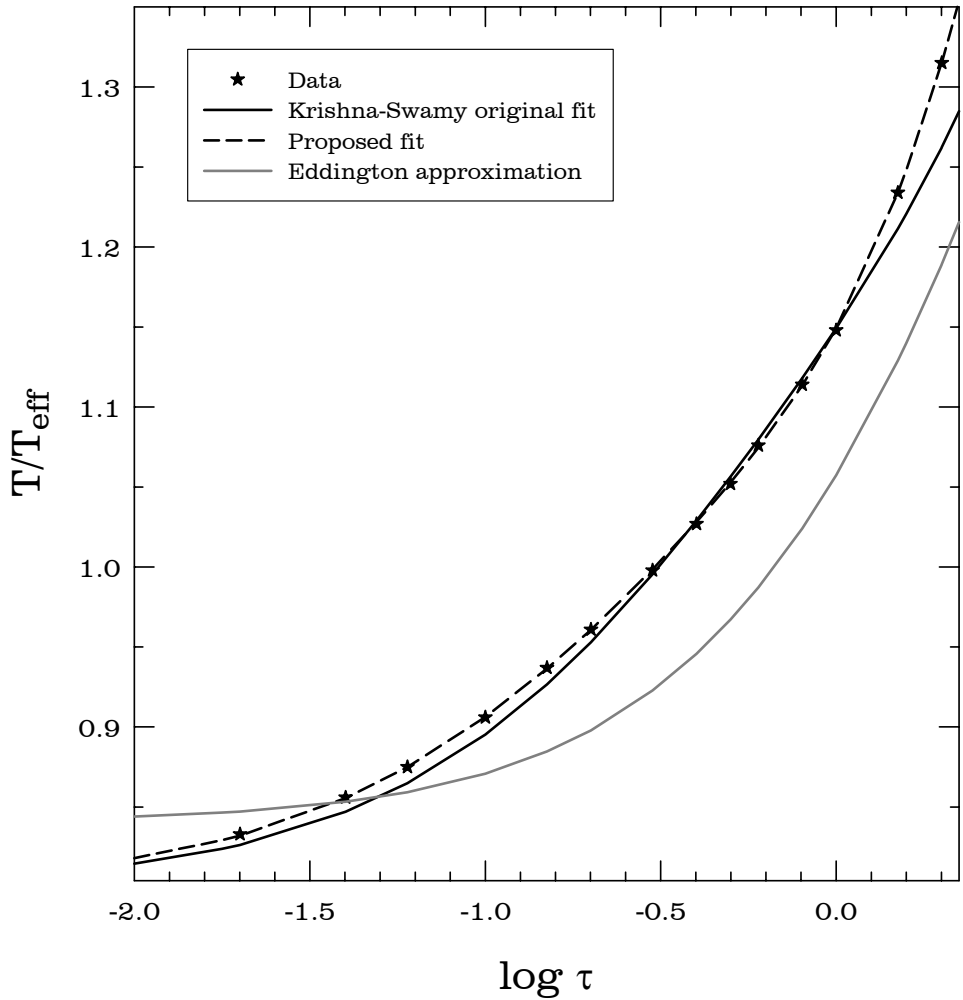}
  \caption{The data originally used in \cite{ks66} to derive the frequently
   employed $T-\tau$ relation ({\em star symbols}) are compared with the original
   Krishna-Swamy fit (eq.~\ref{eq:KRISHNA}, {\em black solid line}) and the new 
   proposed fit (eq.~\ref{eq:KRISNEW}, {\em dashed line}). As can be clearly 
   seen, the new fit represents a significant improvement over the 
   originally suggested one. Also shown is the resulting $T-\tau$ relation
   in the Eddington approximation case {\em gray solid line}).
   }
      \label{fig:newkris}
\end{figure}

While for practical reasons the usage of a semi-empirical $T-\tau$ relation 
such as eq.~(\ref{eq:KRISHNA}) may at present be unavoidable (but see 
\cite{dvea06} for a recent example of evolutionary computations in
which actual model atmospheres were used as boundary conditions), one 
should keep in mind the fundamental limitations of such an approach for 
the production of realistic stellar models. For instance, 
eq.~(\ref{eq:KRISHNA}) is in fact an analytical fit to the data, 
which however detailed inspection reveals not to be very satisfactory: 
as can be seen from Fig.~\ref{fig:newkris} ({\em solid line}), 
where the same data as used to produce  
the original Krishna-Swamy fit is shown as star symbols, the fit provided 
by eq.~(\ref{eq:KRISHNA}) deviates by up to several hundred Kelvin from the 
data in the range between $\tau \simeq 1.5$ and 4, whereas the fit is clearly 
not steep enough for $\tau \lesssim 1.0$. As shown in Fig.~\ref{fig:newkris}
({\em dashed line}), a much improved fit can be obtained, for $\tau \leq 2.0$, 
with the following expression: 

\begin{equation}
T^4 = \frac{3}{4} T_{\rm eff}^4 \left(\tau + 0.55 - 4.46\,\tau^{1.5} + 5.12\,\tau^{2}
      - 1.08\,\tau^{2.5}\right).
\label{eq:KRISNEW}
\end{equation}

In addition, \cite{ks66} warns against using his scaled-solar expression 
for stars of spectral types significantly different from solar, and also when 
convection is efficient as a means of energy transport (as is well known to 
happen, for instance, in RGB and AGB stars). The $T-\tau$ relation for the 
solar atmosphere and for stars of different spectral types is also discussed 
in \S9 of \cite{dg05}.

While the usage of model atmosphere boundary conditions should represent 
a better approach than relying on such $T-\tau$ relations, 
\cite{msea02}, \cite{scea04}, and \cite{dvea06} have indicated that 
major differences do not surface when comparing the results of stellar
models computed with the Krishna-Swamy $T-\tau$ relation and with actual 
model atmospheres. However, the latter should be sufficiently detailed 
that the real physical conditions prevailing in stellar envelopes are 
reproduced; for instance, Fig.~\ref{fig:newkris} reveals that the 
simple Eddington approximation already deviates significantly from 
the semi-empirical $T-\tau$ relations discussed above. 

Some readers may ask, ``why not compute the model atmosphere {\em along} 
with the interior structure?'' While this would be ideal in principle,  
there is a practical limitation that makes this increasingly prohibitive
as the surface is approached. Specifically, when studying the theory of 
stellar atmospheres (see, e.g., \cite{cg68,chea04}), one finds that the 
gradient of the radiation pressure is given by 

\begin{equation}
\frac{dP_{\rm R}}{dr} = -\frac{\kappa_{\rm at}\rho}{c} \frac{L}{4\pi r^2},  
\label{eq:DPRDR}
\end{equation}

\noindent where $\kappa_{\rm at}$ is a {\em flux-weighted opacity} that 
must be computed from 
 
\begin{equation}
\kappa_{\rm at} \equiv \frac{1}{F}\int_{0}^{\infty}\kappa_{\nu}F_{\nu} d\nu,   
\label{eq:KAPPAAT}
\end{equation}

\noindent with $F_{\nu}$ being the monochromatic flux and $F$ the integrated 
net flux. What should be noted here is that $\kappa_{\rm at}$, as given by 
this equation, is {\em not} the same as the Rosseland mean opacity 
$\overline{\kappa_{\rm R}}$ (eq.~\ref{eq:ROSS}). While $\kappa_{\rm at}$ does converge 
to $\overline{\kappa_{\rm R}}$ for large optical depths (see eq.~8.9 in \cite{cg68}), 
this is not the case close to the surface of the star. 
Computing the flux-weighted opacity using eq.~(\ref{eq:KAPPAAT}) 
represents a formidable challenge, since one must know the radiation 
field {\em before} performing the calculations, in addition to having 
all the relevant {\em monochromatic} opacity sources built into the code. 
Unfortunately, this does not look like a realistic approach in the foreseeable 
future.

\subsection{Numerical Techniques for the Integration of the Equations}
\label{sec:numeric}

Once all physical ingredients have been properly taken into account, one 
must choose a numerical technique to carry out the integration of the 
aforementioned equations describing the structure and evolution of 
a star. Two of the most frequently used techniques~-- the 
{\em integration} of {\em fitting method} (e.g., \cite{ms58}) 
and the {\em relaxation} or {\em Henyey method} 
(\cite{lhea64}; see also \cite{ii65})~-- have been described 
in great detail in \cite{ms58,dc68,rkea67}. 
In the former method, the differential equations are independently integrated 
from the inside of the star out, on the one hand, and from the outside in, 
on the other~-- and the solutions are required to match at some intermediate 
point in the stellar interior. One problem with this method is that these 
integrations become extremely sensitive to the starting values 
once stars evolve off the MS (e.g., \cite{ss62}). 
As a consequence, in modern work, the Henyey method is by far the most 
frequently used method. 

Whereas the aforementioned works by \cite{dc68,rkea67} should be consulted
for more details, the overall flavor of the Henyey method can be described
as follows. First, one divides the star into discrete mesh points, then 
replaces the basic differential equations~(\ref{eq:BAS1})-(\ref{eq:BAS4}) 
with difference equations. Given an approximate solution for these 
difference equations, which can come either from a good guess or more
commonly from the previous model in the evolutionary sequence, one then 
linearizes the difference equations to first order in the corrections
and solves the resulting set of linear equations for the corrections,  
which are then added to the approximate solution to obtain an improved 
solution. This process is then repeated until the corrections become 
sufficiently small, i.e., the model converges. In its essence, the 
Henyey method is basically a glorified version of the Newton-Raphson 
method for solving equations given an approximate 
solution. Implementations of the Henyey method are provided in  
the CD that accompanies the \cite{chea04} monograph. 

In stellar evolution work, one is obviously interested in determining 
the {\em time evolution} of the models thus computed. Naturally, one 
must specify a sufficiently small but finite time step to adopt in 
the numerical computations. Such a time step can be obtained by 
requiring that the change in the abundance of the elements
(eq.~\ref{eq:CHEMEVOL}), from  
one model in the sequence to the next, be sufficiently small. However, 
some evolutionary phases of low-mass stars, such as the RGB and AGB, 
present some particular difficulties that prevent such a simple 
procedure to be applied in practice. In particular, during 
these phases one often encounters exceedingly narrow H-burning 
shells, with masses of order $10^{-3}\,M_{\odot}$. 
Consider one such H-burning shell, whose mass coordinate, in the case 
of an RGB star, is evolving outward in time. 
Clearly, even a small outward movement of this shell 
produces a large change in the hydrogen abundance, since the hydrogen 
content above the shell is still high, while that underneath it is 
essentially zero. In terms of the simple method just described, 
then, the time step required to follow 
the time evolution of this RGB star becomes very small, thus requiring 
an enormous number of models, and therefore large amounts of computer 
time, in order to reliably follow the evolution of the star during 
its ascent up the RGB (\cite{ii67a}). For this reason, a special technique 
has been specifically devised and is commonly adopted to treat the time 
evolution of this evolutionary phase. This is the so-called 
{\em shell-shifting} technique, which is described in detail 
in \cite{hs66,ii67,br72,dv92,as94}. 

In like vein, the helium ``flash'' phase which follows the end of 
the RGB phase but precedes the HB  is commonly not followed in 
detail in the course of hydrostatic evolutionary computations (but see 
\cite{tbea01,tlea04} for impressive exceptions to this rule). Instead, 
one commonly applies methods designed to ``transport'' or ``scale'' 
an initial model (taken from the RGB tip) to the zero-age HB phase. 
Different techniques to accomplish this have recently been reviewed by 
\cite{sw05}, to which the reader is referred for a comparison of the 
differences that may result between these approximate techniques and 
actual evolutionary computations carried out through the He flash.

\subsection{Comparison with the Observations}
\label{sec:compobs}

Once a theoretical stellar model has been successfully computed, the main 
output of the code that is of interest to observers (for a star of a given 
mass and chemical composition) 
consists in the total luminosity $L$ and effective temperature 
$T_{\rm eff}$. Photometric observations, on the other hand, do not yield 
directly either of these quantities; instead, one is presented with 
magnitudes (which are converted into {\em absolute} magnitudes once the 
interstellar extinction and distance are known) and {\em color indices}. 
To compare theoretical predictions with the observations, one again needs 
to resort to model atmospheres, from which are computed {\em bolometric 
corrections} and {\em color-temperature relations}. While it is common 
practice to use such relationships derived with a solar-scaled mix of the 
elements, it is well known that stars pertaining to different stellar 
populations may present very different proportions of different types 
of elements, such as (and especially) the $\alpha$-capture elements 
(e.g., \cite{cwea89,bpea05,ccea03,ar06,mzea06} and references therein). 
It has recently been shown 
(\cite{scea04}) that model atmospheres computed with the appropriate 
element abundances may yield appreciably different bolometric 
corrections and color transformations than scaled-solar ones, with a 
potentially significant impact on the comparison between model predictions 
and the observations.

\section{Non-Canonical Effects} 
\label{sec:noncan}

The set of equations in \S\ref{sec:4BAS} is strictly valid only in the 
so-called ``standard'' or ``canonical''  scenario, which 
remains by far the most frequently adopted scenario in the modern 
stellar astrophysics literature. More specifically, in the canonical 
framework one ignores any deviations from spherical symmetry, rotation,   
element diffusion, magnetic fields, and the possible presence of close 
companions, and each and any of the possible associated instabilities. 
Active work on each of these specific areas continues to be carried out, 
but so far a complete evolutionary sequence for a star consistently taking 
into account all of these neglected effects has not been produced~-- and, 
we suspect, is unlikely to be produced in the near future, either. The 
following describes some of the recent work on some of these topics:

\vskip 0.5cm

\noindent {\em 3D Calculations:} 
Interesting results of 3D calculations for selected evolutionary phases, 
especially those 
where hydrodynamical effects are expected to be more important [i.e., 
when the acceleration term that appears in eq.~(\ref{eq:BAS2}) outside 
hydrostatic equilibrium conditions cannot be safely neglected], such 
as the helium flash at the end of RGB evolution, have recently been 
presented (\cite{ddea06}). Encouragingly for the standard stellar 
evolution framework, the final He-burning structures resulting from 
these calculations are not very different from the 1D model predictions. 
Likewise, 3D hydrodynamical calculations appear important to properly 
describe the evolutionary stages preceding the zero-age main sequence 
phase (\cite{wk01}), and (as already stated; see \S\ref{sec:CRIT} above) 
have even been claimed (\cite{peea06}) to provide the key to solve the 
problem of the observed abundances of $^3$He in low-mass stars. 

For model stellar atmospheres, 3D calculations have also 
recently been suggested to be of great importance, leading to fundamental 
changes in quantities such as the chemical composition that are derived 
from spectral observations of low-mass stars such as the Sun (\cite{ma05} 
and references therein). On the other hand, the revised chemical 
abundances suggested in these 3D studies lead to solar models which 
appear to be completely inconsistent with the results of helioseismology
(e.g., \cite{jbea05,sbea07}; see also \cite{jgea06} for a review 
and additional references), which are in much better 
agreement with calculations based on the previously accepted solar 
abundances (\cite{gs98}). This is one of the most important problems 
in stellar structure and evolution at present, and its solution 
may impact several different areas of astrophysics, considering that 
the ``solar metallicity'' is a fundamental point of reference in many 
different astrophysical contexts. 

\vskip 0.5cm

\noindent {\em Rotation:} 
The impact of rotation is one of the major question marks in the study 
of the evolution of low-mass stars. Observations of evolved low-mass 
stars, particularly on the HB phase, reveal a rich phenomenology 
which is very difficult to explain on the basis of simple models 
(see \cite{mc05} for a recent review and references). An illuminating 
description of some of the fundamental problems encountered in treating 
angular momentum transport and evolution in rotating low-mass stars is  
provided in \cite{sp00}, whereas \cite{mc05} contains examples of 
several circumstances under which rotation can play a role in 
interpreting the observed properties of evolved stars in low-mass 
stellar systems. 	

\vskip 0.5cm

\noindent {\em Extra Mixing and Pollution:} 
As is well known, rotation may also induce extra mixing beyond 
what is predicted in the canonical framework, which may be crucial in 
explaining the observed abundance patterns for the light elements in 
low-mass red giants (e.g., 
\cite{sm79,jpz92,cc95,dlrea97,ccb00,dmea00,dnjrea03,dv03,ebv04,apea06}). 
Such extra mixing may not always be due to rotation: tidal effects due 
to the presence of close companions may also play an important role in 
at least some cases (\cite{dh04,pdea06}). 

In regard to abundance inhomogeneities in globular cluster stars, it 
is becoming increasingly clear that a fraction of the stars in at 
least some globular clusters forms from gas that has somehow 
become contaminated by helium-enriched material in the cluster's early 
history, leading to distinct signatures in the observed CMD's 
(e.g., \cite{jn04,lbea04,fdea05,lea05,gpea05,cd07,svea07}). 
However, it remains far 
from clear how the large pollution levels that are inferred from 
these CMD's can come about, though several different scenarios have 
been recently advanced for the origin of these (presumably) 
second-generation stars 
(e.g., \cite{akea06,kbea07,drea07,cy07} and references therein). 
As an alternative to this second-generation/pollution scenario, 
it has also been suggested that protocluster clouds may, as a 
consequence of diffusion, show important helium abundance variations 
from one place to the next \cite{xs95,lc06}. One way or another, the 
confirmation of significant abundance inhomogeneities in at least 
some globular clusters has weakened somewhat the traditional view 
that globular star clusters are the closest approximation to a 
physicist's laboratory in astronomy (\cite{sm01}), a view which is 
generally held due to the great uniformity in their stars' ages, 
distances, and (as had been previously thought) chemical composition.

\vskip 0.5cm

\begin{sidewaysfigure}[!htbp]
  \includegraphics[angle=0,scale=0.825]{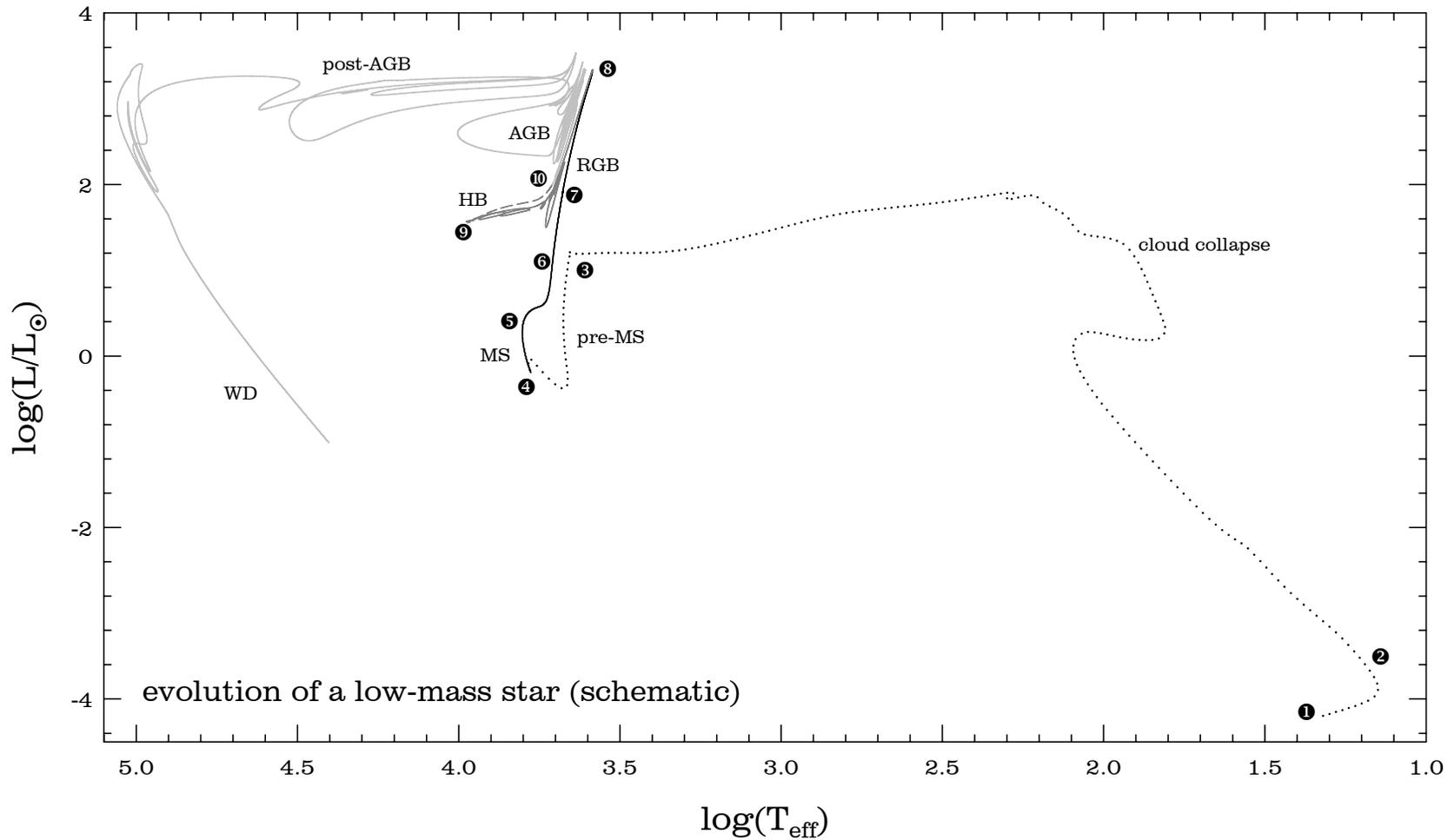}
  \caption{Evolution of a low-mass star in the HRD. All of the main
  evolutionary phases are shown, from its birth during the cloud collapse 
  and pre-MS phase ({\em dotted line}) to its death along the WD cooling 
  curve (hot end of the {\em light gray curve}). The {\em solid black line} 
  indicates the MS and RGB phases, the {\em dark gray line} the pre-HB phase,
  the {\em dashed line} the HB phase, and the {\em light gray line} 
  the AGB, post-AGB, and WD phases. The numbers alongside the evolutionary
  track indicate noteworthy episodes in the life of the star, as discussed
  in \S\ref{sec:evolover}. (The pre-MS and cloud collapse phases are adapted
  from \cite{wk01}, whereas the evolution from the ZAMS onwards has been 
  kindly provided by A. V. Sweigart.) 
  }
  \label{fig:globa-track}
\end{sidewaysfigure}

\noindent {\em Mass Loss:} 
One of the most fundamental, yet least studied, non-canonical ingredients 
in the evolution of low-mass stars is mass loss. It has long been known 
that low-mass RGB stars must lose substantial amounts of mass before 
helium ignition in their cores (e.g., \cite{cr68,ir70,br73,cc93,dcea96}), 
but no fundamental theory exists that 
allows one to compute such mass loss from first principles. While mass 
loss does not affect the evolutionary path of an RGB star in the CMD in 
a substantial way unless mass loss is so substantial that the star may 
either directly become a helium WD or ignite helium on its way to 
the WD cooling curve (e.g., \cite{cc93,tbea01,scea03,tlea04,mcea06}), 
mass loss does have a huge impact on the subsequent evolution, determining 
the temperature distribution along the HB (and therefore playing a key 
role within the framework of the so-called ``second parameter'' phenomenon 
that affects the HB morphology of Galactic globular clusters; see, e.g., 
\cite{mc05} and references therein) and the ultimate fate of the star. 
As to the latter, depending on the total mass loss on the RGB phase, 
the star can go through an  AGB manqu\`e, a post-early-AGB, 
or a post-AGB phase, before finally evolving to the carbon-oxygen 
WD cooling curve (see, e.g., \cite{bdea93}). Last but not 
least, there exist several different semi-empirical mass loss formulae 
that purport to describe the mass loss rates as a function of a star's 
physical parameters (such as radius $R$, surface gravity $g$, and 
luminosity $L$), none of which being clearly superior to the others 
in terms of describing the available data; what is worse, the different 
formulae predict different amounts of mass loss on the RGB, which makes 
it virtually impossible to predict with any degree of confidence the 
temperature that the star will have once it reaches the zero-age HB
(\cite{mc00}; see \S5 in \cite{mc05} for a recent review and 
references). Mass loss may also be a relevant ingredient on the HB
phase proper, at least as far as the derivation of surface gravities 
from the observed hydrogen line profiles is concerned (\cite{vc02}).

\section{The Evolution of a Low-Mass Star: An Overview} 
\label{sec:evolover}

Now that we have built the fundamental equations of stellar structure
and evolution, let us see briefly what has been found to happen, in 
the case of a low-mass star, when these equations are integrated 
numerically and the solution is allowed to evolve with time.

\begin{sidewaysfigure}[!htbp]
  \includegraphics[angle=0,scale=0.825]{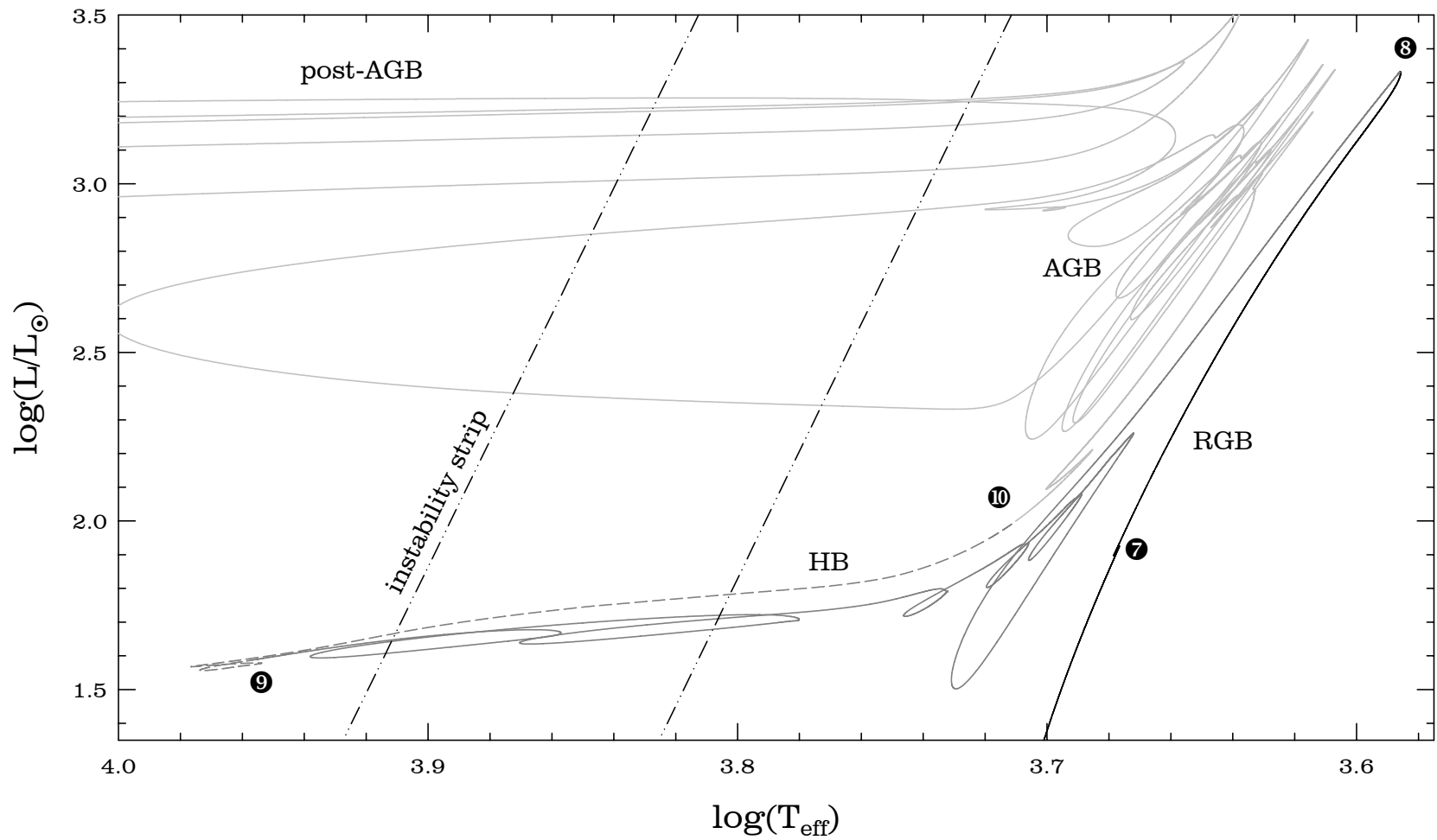}
  \caption{As in Fig.~\ref{fig:globa-track}, but zooming in around the
  HB/upper RGB/AGB phases. The schematic location of the classic
  instability strip is shown as dot-dashed lines.
  }  
  \label{fig:globa-track-zoom}
\end{sidewaysfigure}

In Fig.~\ref{fig:globa-track}, we show the complete evolutionary path 
for one such star. The part of the track labeled ``cloud collapse'' is 
in fact not the result of hydrostatic calculations as have been primarily 
discussed in this paper, but rather represents the results of the 
hydrodynamical simulations performed by \cite{wk01}. All the other 
evolutionary phases, in turn, were kindly computed by A. V. Sweigart 
(2007, {\em priv. comm.}), using his 1D hydrostatic code. 
Note that a small, arbitrary shift (in both temperature and luminosity) 
has been applied to the \cite{wk01} track, in order to ensure a match 
with the Sweigart zero-age main sequence (ZAMS). 

We now provide a brief 
summary of the main aspects of the indicated evolution. More detailed 
discussions can be found in \cite{wk01} and \cite{pb92} (cloud collapse 
phase), \cite{ii65} and \cite{ch66} (hydrostatic contraction to the 
ZAMS), in several papers by Icko Iben, Jr. and collaborators 
(evolution on the MS and beyond; \cite{ii67,ir70,ii82,ii91}), and
also in several monographs (e.g., \cite{dc68,sc05}, etc.). Empirical 
tests of low-mass stellar evolution using the CMD's of globular clusters 
are described in great detail in the review paper by \cite{rfp88}. 
We follow
key phases of the evolution of the (proto-)star according to the numbers 
that are indicated along specified sections of the evolutionary track in 
Figs.~\ref{fig:globa-track} and \ref{fig:globa-track-zoom}: 

\newcounter{local}\renewcommand{\labelenumi}
  {\setcounter{local}{181+\value{enumi}}%
  \ding{\value{local}}}
\begin{enumerate}
 \item We begin by considering a molecular cloud that becomes unstable 
    according to the Jeans (\cite{jj02}) criterion, and which accordingly 
    collapses and fragments in the process (see \cite{pb92} for a review). 
    Under normal conditions, some of these fragments~-- most, as a matter 
    of fact~-- will eventually give rise to low-mass stars. At this early 
    phase (i.e., prior to \ding{182}), however, the temperature of the fragment 
    is still very low, of order 10~K, and it cannot rise much because the 
    cloud is still optically thin. Accordingly, this phase~-- not shown 
    in Fig.~\ref{fig:globa-track}~-- is known as the {\em isothermal 
    phase} of the cloud collapse. Eventually the opacity will increase 
    and the cloud will become optically thick, thus departing from 
    isothermality. When this happens, the proto-star has reached point
    \ding{182} in Fig.~\ref{fig:globa-track}, and the classical ``adiabatic
    phase'' commences. According to the calculations provided in \cite{wk01}, 
    it takes the proto-star of order $1.5\times 10^5$~yr to reach this point. 

 \item Around this point a hydrostatic core forms for the first time, and the 
    main accretion phase starts. As can be seen, the luminosity and temperature 
    increase steadily, and the irregularities that are seen in the evolutionary 
    track reflect random fluctuations in the mass accretion rate. 

 \item When accretion tapers off and the final mass is approached, the 
    photosphere of the proto-star finally becomes visible, the structure 
    becomes fully convective, and its luminosity decreases at almost constant 
    temperature, leading to the vertical segment of evolutionary track~-- the 
    so-called {\em Hayashi track} (representing the canonical pre-MS phase)~-- 
    shown in Fig.~\ref{fig:globa-track}. Along most of this vertical segment, 
    the proto-star is fully convective, and the core gets sufficiently 
    hot that deuterium is burnt efficiently. A radiative core only 
    forms at the bottom of the curve, due to the fact that the increasing 
    core temperature also leads to increased ionization and hence to a 
    decrease in the opacity, which as we have seen tends to quench convection 
    (recall eqs.~\ref{eq:SCHWARZ} and \ref{eq:DELRAD}). After this point,   
    the proto-star becomes hotter and increases again in brightness while 
    it is still shrinking in size. The increasing core temperatures also 
    lead to incomplete CNO processing along the first steps of the 
    CN-branch of the CNO cycle, with $^{12}$C in particular being 
    progressively consumed in the core. (For even lower-mass stars, 
    the core temperature never becomes high enough for $^{12}$C to be 
    processed, and the pre-MS evolution remains vertical all the way to 
    the ZAMS.) Until $^{12}$C is finally
    exhausted in the core and the PPI chain finally begins to operate 
    in equilibrium, with the star settling on the ZAMS, 
    an additional $5\times 10^7$~yr will have passed (\cite{ii65}). 
 
\item This is the ZAMS proper. When the star finally settles here, it stops 
    contracting and now evolves along the much longer {\em nuclear timescale}, 
    which is of order $10^{10}$~yr for low-mass stars. Accordingly, the star 
    changes position along this region of the HRD only very slowly, with its
    temperature and luminosity slightly increasing with time as hydrogen is 
    steadily transformed into helium by means of the PP chain (the CNO
    cycle is currently responsible for a mere 1.5\% of the Sun's current
    luminosity; see \cite{jbea01}). The MS phase
    is so slow that the Sun's radius had already attained about 87\% 
    of its current value, its effective temperature was around 97\% of 
    the current temperature, and its luminosity some 68\% of the current 
    luminosity when the Sun first reached the ZAMS (\cite{jbea01}), which 
    happened some $4.57\times 10^9$~yr ago (\cite{bp95,ms02}). A very nice 
    pie chart illustrating the run of several physical quantities with 
    radius for a low-mass ZAMS star is provided in Fig.~7 of \cite{ii71}.
  
\item This is the so-called {\em turn-off point}, which is related to 
    (though not necessarily identical with) the exhaustion of hydrogen 
    in the center of the star. From this point onwards, hydrogen burning 
    ceases to be a central process, becoming instead a {\em shell-burning}
    process. Initially, hydrogen burns in a thick shell, whereas the 
    helium-rich core, which is now becoming isothermal, keeps growing 
    in size as a consequence of the helium that is being produced at its 
    outer border. The core cannot grow indefinitely, however: there is
    a limit to the mass that can exist in an isothermal core and still
    support the overlying layers, given by the so-called 
    {\em Sch\"onberg-Chandrasekhar mass} (\cite{sc42}), or around 10\% 
    of the total mass of the star. When its mass increases beyond this
    point, the core must collapse on a thermal timescale, thus heating 
    up and releasing energy in the process. The 
    rise in core temperatures also leads to a rise in temperature at 
    the base of the thickening H-burning shell, and accordingly   
    H-burning by the CNO cycle becomes increasingly important (having 
    in fact become dominant in the center even before the TO point 
    was reached). As a 
    consequence of the fact that the CNO cycle has a very high temperature
    dependence, the H-burning shell becomes thinner and thinner. However, 
    not all the energy released by the H-burning shell actually reaches
    the surface: part of it is used to expand 
    the star's envelope. As a consequence of this expansion, the star
    begins to cool down, and the characteristic evolution to the right 
    constitutes the so-called {\em subgiant phase}. (In higher-mass stars, 
    this phase proceeds much faster, and gives rise to the so-called 
    {\em Hertzsprung gap} in the color-magnitude diagram.) The 
    corresponding cooling of the outer layers leads to the formation of
    a convective envelope, and the star eventually reaches the base of 
    the RGB. The further contraction of the core, moreover, leads to the 
    establishment of electron degeneracy there. Therefore, when  
    the star reaches the base of the RGB, its structure is characterized 
    by a growing but inert, partially degenerate He-core, surrounded by 
    a H-burning shell that is becoming progressively thinner, and an 
    outer convective envelope that is becoming progressively thicker. 
    
\item At this point, the convective envelope reaches its maximum inward
    penetration, thus leading to the dredge up of material that has 
    been partially processed nuclearly, including a small amount of 
    He. This is the so-called {\em first dredge-up} phase (\cite{ii64}). 
    From this point in time onwards, the convective envelope begins to 
    retreat. It takes the star an additional $10^9$~yr to go from the 
    turn-off point \ding{186} to this evolutionary stage, whereas the
    remainder of the RGB phase lasts of order $10^8$~yr, becoming
    progressively quicker as higher and higher luminosities are
    attained.  
    
\item The H-burning shell continues to advance outward in mass, thus
    leading to a continued increase in mass of the He core. At the 
    indicated point, the H-burning shell actually encounters the 
    chemical composition discontinuity that was left behind as a 
    consequence of the maximum inward penetration of the convective
    envelope. Since the envelope is naturally H-rich, this means that 
    the H-burning shell is suddenly presented with an extra supply of
    fuel. The structure of the star, presented with this extra fuel
    supply, readjusts momentarily to this new situation, with an 
    actual (small) reversal in its direction of evolution before
    it resumes its ascent of the RGB. The details of this process 
    depend crucially on the precise abundance 
    profile in the H-burning shell (see, e.g., \cite{scea02}). 
    In the observed CMD's and RGB luminosity functions of globular 
    star clusters, and as first predicted by \cite{hct67,ii68},     
    one in fact identifies the so-called {\em RGB ``bump''} as an 
    observed counterpart of this stellar interior phenomenon (e.g., 
    \cite{ckea85,ffpea90}). Importantly, the RGB bump also appears 
    to correspond to the position in the CMD that marks the onset of
    mixing of nuclearly-processed elements beyond that predicted by 
    the canonical theory (e.g., \cite{rgea00,dv03,cc05,rbdl07} and 
    references therein).    
    
\item The final stages of the star's ascent of the RGB are again 
    characterized by the increasing size of the He-core, which keeps
    on contracting and heating up. Very instructive pie charts showing
    schematically the structure of the envelope and core of such an 
    RGB star have been provided, respectively, in Figs.~8 and 9 of
    \cite{ii71}. At this stage large amounts of 
    energy are lost in the form of neutrinos, such emission being 
    most efficient where the matter is denser~-- which leads to an 
    actual temperature {\em inversion} in the core 
    (\cite{hct67,ii68a,br72,sg78}), its hottest regions moving as 
    far away from the center as $\sim 0.25\,M_{\odot}$ by the time 
    the RGB tip is reached. 
    As a consequence, when temperatures finally become high
    enough for the helium-burning reactions to commence, this happens
    not in the actual center of the star, but in fact in a {\em shell} 
    inside the He-rich core. In this sense, the fact that the 
    matter in the core is degenerate leads to a dramatic phenomenon
    which had not been previously encountered in the life of the 
    star~-- the so-called helium {\em ``flash''}.
    
    \hskip 0.5cm To understand the physics behind the flash phenomenon 
    (see \cite{sh62}), 
    consider first what happens in the case of thermonuclear burning under 
    {\em non}-degenerate conditions. In this case, the energy input
    to the medium by the nuclear reactions tends to increase the 
    local temperature, but this temperature increase also tends to 
    increase the local pressure, which in turn tends to expand the
    region where the reaction is taking place and hence cool down
    the material, thereby preventing a further increase in the
    nuclear reaction rate. In other words, thermonuclear burning, under 
    non-degenerate conditions, is a self-regulating process, which
    allows nuclear burning to proceed quiescently over long
    periods of time. In the case of degenerate matter, on the other 
    hand, this is not possible: since in this case the EOS lacks a 
    temperature dependence, the local temperature increase does 
    {\em not} lead to an increase in the local pressure, and 
    therefore the region where burning has started cannot expand
    and cool down. As a consequence, the energy input by the 
    nuclear reactions leads to an increase in the local temperature,
    which leads to a further increase in the thermonuclear reaction
    rates, which leads to another increase in the local temperature,
    and so on and so forth~-- giving rise to a so-called 
    {\em thermonuclear runaway} (hence the term ``He flash''), which 
    can only cease once degeneracy is lifted. As a matter of fact, 
    most of the energy that is produced during the He flash (which 
    can exceed, according to \cite{sh62}, 
    $10^{12}\,L_{\odot}$; in fact,~the models shown in 
    Figs.~\ref{fig:globa-track} and \ref{fig:globa-track-zoom} 
    reach a peak He-burning luminosity of $9.2\times 10^{9}\,L_{\odot}$, 
    which should be compared with the luminosity emitted by the entire
    Galaxy, $\sim10^{10}\,L_{\odot}$) 
    is used to lift up the degeneracy in the core, so that the 
    luminosity of the star does not increase during this so-called
    ``pre-HB phase''~-- rather the opposite, in fact (see 
    Fig.~\ref{fig:globa-track-zoom}). In addition, since (as already 
    stated) the initial flash occurs off-center, additional flash
    events take place increasingly closer to the center of the star,
    until degeneracy has been lifted throughout the He core
    (\cite{ms81,as94b}). When
    this happens and the star is finally able to burn helium 
    quiescently in a convective core and hydrogen in a shell, the
    star has reached the so-called zero-age HB (ZAHB). From
    the RGB tip to the ZAHB, $\sim 10^6-10^7$~yr will have passed, 
    and around 5\% of the core's helium fuel will have been consumed
    (\cite{as94b}). 
    For this reason, only a very small number of stars is expected
    to be found in the pre-HB evolutionary phase, which is the 
    reason why no such star has ever been positively identified.
    Very recently, it has been suggested that, as a consequence
    of the fact that many pre-HB stars are expected to cross the 
    instability strip before they arrive on the ZAHB, some stars
    which might be going through this phase could be identified  
    as variable stars with peculiarly fast period change rates 
    (\cite{mc05}), since their periods should depend, according
    to the period-mean density relation of pulsation theory, on 
    their luminosities and radii, which are both changing on a 
    very short timescale compared with other variables (such as 
    the RR Lyrae and type II Cepheids) which occupy a similar region
    of the HRD.    
    
\item This is the ZAHB that was alluded to in the previous paragraph,
    marking the onset of the He-burning phase proper~-- which, 
    in the case of low-mass stars, is referred to as the 
    {\em horizontal branch} (HB) phase, due to the fact that 
    HB stars all have very nearly the same luminosity irrespective
    of mass, thus leading to the characteristic horizontal feature
    that is seen in the CMD's of globular clusters. HB stars play
    a particularly important role in determining the distances and 
    ages of old stellar populations, and also in the interpretation 
    of the formation and evolution history of the Galaxy and its
    neighboring satellites (see \cite{mc05} for a recent review).  
    Pie charts illustrating the run of several physical quantities with 
    radius for a ZAHB star located close to the blue edge of the 
    instability strip are provided in Figs.~10 and 11 of \cite{ii71}.
    
    \hskip 0.5cm Unfortunately, the exact ZAHB temperature of a star cannot 
    be predicted a priori, depending as it does on how much mass the 
    star may have lost on its ascent of the RGB (see the end of 
    \S\ref{sec:noncan} for a discussion of some of the problems 
    encountered in treating mass loss in red giant stars). In general, 
    the stars that lose the least amount of mass will fall on the red
    portion of the ZAHB, whereas those that lose a substantial fraction
    of their masses will actually fall on the blue ZAHB, some possibly 
    becoming ``extreme'' HB (EHB) or blue subdwarf (sdB) stars. In 
    turn, intermediate amounts of mass loss should lead to RR Lyrae
    variable stars, since the classical instability strip 
    actually crosses the HB at intermediate temperatures, as can be 
    inferred from Fig.~\ref{fig:globa-track-zoom}. Of course, not only
    stars whose ZAHB positions fall within the instability strip
    will become RR Lyrae variables: in fact, as a consequence of 
    evolution away from the ZAHB, stars which arrived on the blue
    ZAHB are expected to become RR Lyrae stars later in their lives, 
    when they are already well on their way to become AGB stars
    (which is precisely what happens with the particular star whose
    evolutionary path is shown in Fig.~\ref{fig:globa-track-zoom}). 
    Conversely, some stars whose ZAHB position falls on the red HB
    may also, depending (among other things) on the relative 
    efficiency of their H-burning shells compared with their He-burning
    cores, develop blueward loops that may take them through the 
    instability strip at some point in the course of their lifetimes. 
    
\item When the star reaches this point, which happens $\sim 10^8$~yr
    after its arrival on the ZAHB, He has been exhausted in its
    center, and the star begins its life as a low-mass AGB star. 
    A glimpse at Fig.~\ref{fig:globa-track-zoom} will promptly reveal
    that the term ``asymptotic'' comes from the fact that, for low-mass 
    stars, the evolution
    proceeds ever closer to the first-ascent RGB, but never quite 
    reaching it. In terms of its internal structure, an AGB star  
    is defined as a star with an inert core (comprised mostly 
    of carbon and oxygen) which burns He in a shell and H in another 
    shell further out, such shells becoming progressively thinner as 
    the star climbs up the AGB. A very instructive diagram comparing 
    the run with radius of several different physical quantities for 
    an HB and an AGB star is provided in Fig.~4 of \cite{ir70}. 
    
    \hskip 0.5cm In AGB stars, the He- and H-burning shells take 
    turns as the most efficient energy sources, as a consequence of the 
    thermal instabilities that were first encountered in the 
    computations by \cite{sh65}. 
    In spite of being reminiscent of the He flashes that 
    were found under degenerate conditions at the RGB tip and on the 
    pre-HB phase, electron degeneracy does not play an important role 
    in the case of these so-called ``AGB thermal pulses,''     
    the cause here being instead a highly temperature-dependent
    nuclear energy source operating in a very thin shell (\cite{sh65}). 
    Stars that have 
    not yet reached this thermally pulsing phase are termed ``early AGB'' 
    or E-AGB stars (\cite{ir83}), whereas the thermally pulsing ones, 
    which are preferentially found towards the brighter portions of 
    the AGB, are called, not surprisingly, TP-AGB stars. 

    \hskip 0.5cm In E-AGB stars,
    the H-burning shell has very little efficiency, since the onset of
    He burning in a shell has caused the H-burning shell to expand and 
    cool. The onset of He-shell burning causes a temporary reversal in 
    the star's evolutionary direction (as can be seen immediately after
    point \ding{191} in the evolutionary track shown in 
    Fig.~\ref{fig:globa-track-zoom}), which leads to the so-called 
    {\em AGB clump} in the observed CMD's of well-populated globular 
    clusters and old Local Group galaxies (e.g., \cite{lp92,cg98,scea01}).
    Since the H-burning shell is inefficient, the mass of the 
    He-rich region outside the C-O core cannot increase much 
    in mass during this phase. Eventually, the outward-moving 
    He-burning shell 
    reaches the associated He/H chemical composition discontinuity.      
    When this happens, the He-burning shell dies down, and 
    accordingly the overlying H-burning shell contracts rapidly and
    is re-ignited. The corresponding He ashes are then compressed and 
    heated, and finally ignite once they reach a critical value (of 
    order $10^{-3}\,M_{\odot}$ for a $0.8\,M_{\odot}$ C-O core; see
    \cite{sc05}). This marks the onset of the 
    TP-AGB phase. 

    \hskip 0.5cm Unlike what happened in the case of
    the earlier He flash at the tip of the RGB, matter in this case is
    {\em not} degenerate, so that the energy that is liberated in a 
    pulse does lead to an increase in the local pressure, and 
    therefore to an expansion. 
    This expansion cools down the region where 
    He burning is taking place, thus lowering its rate and
    leading to a phase of quiescent He burning. The H-burning shell, 
    during this expansion phase, is pushed outward to such   
    low temperatures that H burning again dies down. 
    During this quiescent He-burning phase, the 
    He-burning shell consumes all the He that was created by the 
    previous phase of H burning in a shell, until a  
    He/H discontinuity region is again reached. As before, 
    He burning is then again quenched, H burning resumes, and
    the same processes that had previously given rise to a thermal 
    pulse are again triggered.     

    \hskip 0.5cm Besides nucleosynthetic signatures in the form of 
    $s$-process elements (see \S\ref{sec:unc}), these thermal pulses 
    can give rise to characteristic ``loops'' in the CMD, such as the 
    ones that can be seen in Fig.~\ref{fig:globa-track-zoom} (see, 
    e.g., \cite{sh70,rg74,ds79,ii82}). During some of these loops, the 
    star may again cross the instability strip, becoming a type II
    Cepheid (\cite{sh70,rg76}). The details of the AGB evolutionary 
    phase, which lasts of order $10^7$~yr, are extremely 
    complex, depending strongly on the (poorly known) mass loss rates, 
    and the reader is referred to the quoted papers, and also to the
    recent monograph by \cite{ho04} and reviews by \cite{ir83,fh05,sc05}, 
    for more detailed descriptions of the physics and phenomenology 
    surrounding AGB stars. 
    
    \hskip 0.5cm Once the AGB star's envelope mass has become very low, a 
    final, quite dramatic mass ejection episode may take place, the so-called 
    ``superwind phase'' (\cite{ar81}), whose origin and nature still remain
    the subject of much debate. One way or another, the ejection of the
    outer layers of the AGB stars is expected to lead to the formation of 
    a so-called ``post-AGB star'' (see Fig.~\ref{fig:globa-track-zoom}), 
    which is basically the bare core of the progenitor AGB star finally 
    exposed. (Surrounding such a post-AGB star a planetary nebula may 
    also be found, in which the ejected gas is being ``lit up'' by the 
    central star.) Following a very quick evolution to the blue accross 
    the HRD, the star finally settles on the WD cooling sequence 
    (Fig.~\ref{fig:globa-track}). The latter, in fact, 
    represents the ``death bed'' of a low-mass star, provided, of
    course, it does not somehow acquire mass from/merge with a 
    companion~-- but that would be quite a different story. 
    
    \hskip 0.5cm Before closing, we note
    that a low-mass star may reach the WD cooling curve following 
    other evolutionary routes, including the following: (i)~Directly
    after the RGB, bypassing the HB and AGB phases, which may happen
    if somehow mass loss on the RGB is so extreme that the He core
    never gets to grow to the minimum level required for He ignition.
    In this case, a He WD is formed, not a C-O one. (ii)~Directly 
    after the HB, bypassing the AGB phase, which may happen if the 
    star arrives on the ZAHB at a very high temperature (i.e., as
    an EHB or sdB star). This has been called the ``AGB manqu\`{e}''
    route (\cite{gre90}). (iii)~Directly from the (early) AGB phase, 
    i.e., prior to the thermal pulsing phase. This so-called 
    ``post-early AGB'' route (\cite{ebea90}) can take place also in
    the case of EHB or sdB stars, though for cooler ones than in the 
    case of AGB manqu\`{e} stars (see, e.g., \cite{bdea93}). Post-AGB,
    AGB manqu\`{e}, and post-early AGB stars, when present, can be
    important contributors to the integrated ultraviolet light from
    old stellar populations (see, e.g., 
    \cite{kdb85,bdea93,bdea95,smea98,ro99}). 

\end{enumerate}


\begin{theacknowledgments}
  I would like to thank the organizers for the invitation and for helping 
  make our stay in Rio so pleasant. I am particularly indebted to A. V. Sweigart 
  for providing some of the data (MS evolution and later) that was used in 
  producing Figs.~\ref{fig:globa-track} and \ref{fig:globa-track-zoom}, 
  and to him and R. T. Rood for several perceptive comments and useful 
  suggestions. A critical reading of the manuscript by H. A. Smith and 
  A. Valcarce is also gratefully acknowledged. Financial support 
  was provided by the organizing committee and by Proyectos FONDECYT 
  Regulares No. 1030954 and 1071002. 
\end{theacknowledgments}

\bibliographystyle{aipproc}   

\end{document}